%% file: main.tex
\pdfoutput=1
\documentclass[12pt,a4paper]{article}

\usepackage{graphicx}  

\usepackage{xspace}
\usepackage{color}
\usepackage{colortbl}
\usepackage{subfigure}
\usepackage{relsize}
\usepackage{sidecap}
\usepackage{color}
\usepackage{colortbl}
\usepackage{afterpage}
\usepackage{amsmath} 
\usepackage{ifthen} 
\graphicspath{{./figures/}} 

\newboolean{pdflatex}
\setboolean{pdflatex}{true} 
\usepackage{rotating} 
\newboolean{articletitles}
\setboolean{articletitles}{true} 

\newboolean{uprightparticles}
\setboolean{uprightparticles}{false} 
\usepackage{amsmath} 
\usepackage{amssymb}
\usepackage{amsfonts}
\usepackage{upgreek} 

\input{lhcb-symbols-def} 

\usepackage{hyperref}    
\usepackage[all]{hypcap} 

\usepackage{bm}
\usepackage{afterpage}

\def\epem     {\ensuremath{e^+e^-}\xspace}
\def\DJ     {\ensuremath{\D_J}\xspace}
\def\DstarPi {\ensuremath{\Dstarp \pim}\xspace}
\def\DPi {\ensuremath{\Dp \pim}\xspace}
\def\DzPi {\ensuremath{\Dz \pip}\xspace}
\def\DTwentyFourThirty {\ensuremath{{D}^\prime_1(2430)}\xspace}
\def\DTwentyFourTwenty {\ensuremath{D_1(2420)}\xspace}
\def\DTwentyFourTwentyNeutral {\ensuremath{ D_1(2420)^0}\xspace}
\def\DTwentyFourTwentyCharged {\ensuremath{ D_1(2420)^+}\xspace}
\def\DTwentyFourSixty {\ensuremath{D^*_2(2460)}\xspace}
\def\DTwentyFourSixtyNeutral {\ensuremath{ {D}^*_2(2460)^0}\xspace}
\def\DTwentyFourSixtyCharged {\ensuremath{ {D}^*_2(2460)^+}\xspace}
\def\DTwentyFourHundred {\ensuremath{D^*_0(2400)}\xspace}

\def\DTwentyFiveFiftyNeutral {\ensuremath{D_{J}(2580)^0}\xspace}

\def\DTwentySixHundred {\ensuremath{D^*_{J}(2650)}\xspace}
\def\DTwentySixHundredNeutral {\ensuremath{D^*_{J}(2650)^0}\xspace}

\def\DTwentySevenFiftyNeutral {\ensuremath{D_{J}(2740)^0}\xspace}

\def\DTwentySevenSixty {\ensuremath{D^*_{J}(2760)}\xspace}
\def\DTwentySevenSixtyNeutral {\ensuremath{D^*_{J}(2760)^0}\xspace}
\def\DTwentySevenSixtyCharged {\ensuremath{D^*_{J}(2760)^+}\xspace}

\def\DThreeNeutral {\ensuremath{D^*_{J}(3000)^0}\xspace}
\def\DThreeCharged {\ensuremath{D^*_{J}(3000)^+}\xspace}
\def\DThreeU {\ensuremath{D_{J}(3000)^0}\xspace}
\def\invfb   {\ensuremath{\mbox{\,fb}^{-1}}\xspace}

\def\mthetah     {\mbox{$\theta_{\rm H}$}\xspace}

\def\cthetah     {\mbox{$\cos\theta_{\rm H}$}\xspace}
\def\DTwentyFiveFiftyNeutralb {\ensuremath{{D}(2550)^0}\xspace}

\def\DTwentySixHundredNeutralb {\ensuremath{{D^*}(2600)^0}\xspace}
\def\DTwentySixHundredChargedb {\ensuremath{{D^*}(2600)^+}\xspace}
\def\DTwentySevenFiftyNeutralb {\ensuremath{{D}(2750)^0}\xspace}

\def\DTwentySevenSixtyNeutralb {\ensuremath{{D^*}(2760)^0}\xspace}
\def\DTwentySevenSixtyChargedb {\ensuremath{{D^*}(2760)^+}\xspace}
\usepackage{cite} 
\usepackage{mciteplus}
\usepackage{amsbsy} 
\begin{document} 

\renewcommand{\thefootnote}{\fnsymbol{footnote}}
\setcounter{footnote}{1}

\input{title-LHCb-PAPER}

\renewcommand{\thefootnote}{\arabic{footnote}}
\setcounter{footnote}{0}

\pagestyle{plain} 
\setcounter{page}{1}
\pagenumbering{arabic}


\section{Introduction}
\label{sec:Introduction}
\begin{figure}[b]
  \begin{center}
    \hspace*{0.1cm}\includegraphics[width=0.95\linewidth]{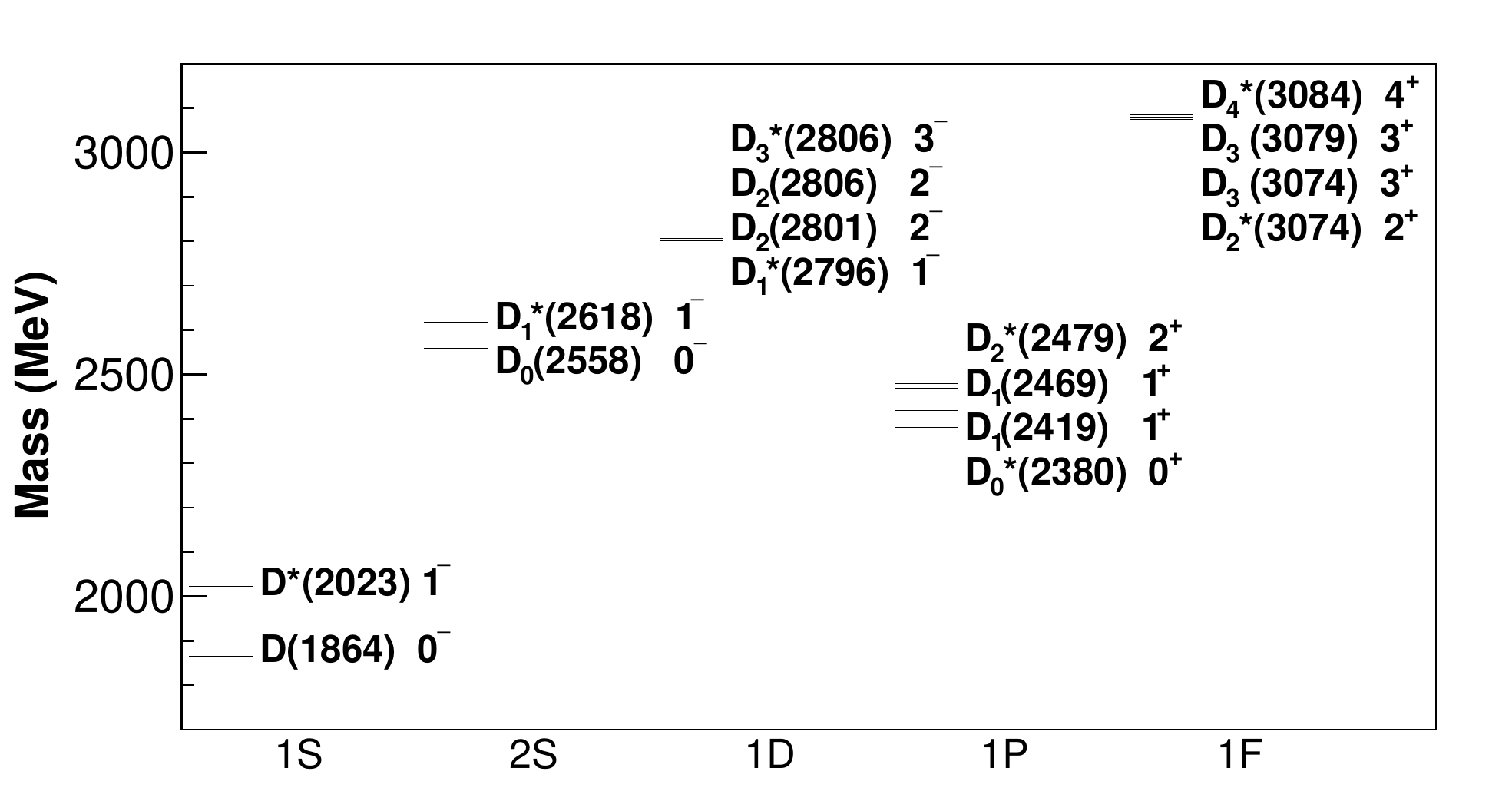}
    \vspace*{-1.0cm}
  \end{center}
  \caption{\small Modified Godfrey-Isgur mass predictions~\cite{Godfrey:1985xj}. The figure shows the $c\bar{u}$ spectrum 
  in which the masses have been scaled such that the ground state coincides with the $D^0$ mass. 
  The $2^-$ states, not shown in the original publication, have been inserted following 
  the splitting structure of the 1P states.
    }
  \label{fig:Godfrey}
\end{figure}

Charm meson spectroscopy provides a powerful test of the quark model predictions of the Standard Model.
Many charm meson states, predicted in the 1980s~\cite{Godfrey:1985xj}, have not yet been observed experimentally. 
The expected spectrum for the $c\bar{u}$ system is shown in Fig.~\ref{fig:Godfrey} (the spectrum of the $c\bar{d}$ system is almost identical). The $J^P$ states having $P=(-1)^J$ and therefore $J^P=0^+,1^-,2^+,...$ are called 
natural parity states and are labelled as $D^*$, while unnatural parity indicates the series $J^P=0^-,1^+,2^-,...$. 
The low-mass spectrum of the $c\bar{u}$ system is comprised of the ground states (1S), the orbital excitations with angular momentum $L$=1,\ 2 (1P,\ 1D), and the first radial excitations (2S). 
Apart from the ground states ($D,D^*$), only two of the 1P states, \DTwentyFourTwenty and \DTwentyFourSixty~\cite{Beringer:1900zz}, are experimentally well established since they have relatively narrow widths ($\sim$30\mev).~\footnote{We work in units where $c = 1$.} In contrast, the broad $L=1$ states, \DTwentyFourHundred and \DTwentyFourThirty, have been established by the \belle and \babar \ experiments in exclusive \B decays~\cite{Abe:2003zm,Aubert:2009wg}.

The theoretical predictions are in agreement (within 20--30~MeV) with observations for the 1S states and the $J^P=2^+$ 
and $J^P=1^+$ 1P states. 
In the $c \bar s$ system, the $J^P=0^+$ and $J^P=1^+$ states (both $L=1$) have
predicted masses about 100~\mev higher 
than the measured masses of the $D_{sJ}$ mesons.
To quantitatively assess the accuracy of the quark model predictions, assumptions are
needed to formulate a wave equation for quark-antiquark bound states starting from the QCD Lagrangian~\cite{Colangelo:2005gb}. 
Nevertheless, the discrepancy between 
the predictions of various models and the mass measurements~\cite{DiPierro:2001uu,Ebert:2009ua,Colangelo:2012xi,Colangelo:2000dp} suggests that some observed 
states in the $c \bar s$ case are not simple quark-antiquark configurations. Possible interpretations include more complex structures, 
such as bound states 
(``molecules'') of other mesons~\cite{Barnes:2003dj}, or mixtures of conventional quark-antiquark with 
four-quark components~\cite{Vijande:2006hj}.

The properties of hadrons  can be computed from the QCD Lagrangian using lattice calculations and
the resulting  
$c \bar u$, $c \bar d$ and $c \bar s$  mass spectra can be compared to measurements. However, the calculation of hadronic quantities for 
dynamical light quarks is still a challenging task and different results are obtained~\cite{Bali:2003jv,Dougall:2003hv,Lin:2006vc,Dong:2009wk,Gong:2011nr,Mohler:2012na}.

To search for excited charmed mesons, labelled \DJ, \babar \ analyzed 
the inclusive production of the \DPi, \DzPi and \DstarPi final states in 
the inclusive reaction ${\mbox \epem \rightarrow c \bar c \rightarrow D^{(*)}\pi X}$, where 
$X$ is any additional system~\cite{delAmoSanchez:2010vq}.~\footnote{Throughout the paper use of charge-conjugate decay modes is implied.} 
They observe four signals, labelled \DTwentyFiveFiftyNeutralb, \DTwentySixHundredNeutralb, \DTwentySevenFiftyNeutralb and \DTwentySevenSixtyNeutralb, and the isospin partners \DTwentySixHundredChargedb and \DTwentySevenSixtyChargedb. 

This paper reports a search for \DJ mesons in a data sample, corresponding to an integrated luminosity of 1.0\invfb, of $pp$ collisions collected at a centre-of-mass energy of $7\tev$ with the \lhcb detector.

\section{Detector}
\label{sec:Detector}

The \lhcb detector~\cite{Alves:2008zz} is a single-arm forward
spectrometer covering the \mbox{pseudorapidity} range $2<\eta <5$, designed
for the study of particles containing \bquark or \cquark quarks. The
detector includes a high precision tracking system consisting of a
silicon-strip vertex detector surrounding the $pp$ interaction region,
a large-area silicon-strip detector located upstream of a dipole
magnet with a bending power of about $4{\rm\,Tm}$, and three stations
of silicon-strip detectors and straw drift tubes placed
downstream. The combined tracking system has momentum resolution
 that varies from 0.4\% at 5\gev to 0.6\% at 100\gev,
and impact parameter resolution of 20\mum for tracks with high transverse momentum
\pt with respect to the beam direction. 
The impact parameter is defined as the perpendicular distance between the track path and the position of a $pp$ collision.
Charged hadrons are identified using two
ring-imaging Cherenkov (RICH) detectors. Photon, electron and hadron
candidates are identified by a calorimeter system consisting of
scintillating-pad and pre-shower detectors, an electromagnetic
calorimeter and a hadronic calorimeter. Muons are identified by a 
system composed of alternating layers of iron and multiwire
proportional chambers. The trigger~\cite{Aaij:2012me} consists of a hardware stage, based
on information from the calorimeter and muon systems, followed by a
software stage which applies a full event reconstruction. 

\section{Event selection}
\label{sec:Selection}

The search for \DJ mesons is performed using the inclusive reactions
\begin{equation}
pp \to \Dp \pim X, \ pp \to \Dz \pip X , \ pp \to \Dstarp \pim X,
\end{equation}
where $X$ represents a system composed of any collection of charged and neutral particles.

The charmed mesons in the final state are reconstructed in the decay modes ${\mbox \Dp \to K^-\pi^+\pi^+}$, $\Dz \to \Km \pip$ and $\Dstarp \to \Dz \pip$.
Charged tracks are required to have good track fit quality, momentum $p>3\gev$ and $\pt>250\mev$. 
These conditions are relaxed to lower limits for the pion originating directly from the \Dstarp decay. Tracks pointing to a $pp$ collision vertex (primary vertex) are rejected by means of an impact parameter requirement in the reconstruction of the \Dp and \Dz candidates. All tracks used to reconstruct the mesons must have a distance of closest approach to each other smaller than 0.5\mm.
The cosine of the angle between the momentum of the $D$ meson candidate and its 
direction, defined by the positions of the primary vertex and the meson decay vertex, is required to be larger than 0.99999. This ensures that the $D$ meson candidates are produced at the 
primary vertex and reduces the contribution from particles originating from $b$-hadron decays.
 \afterpage{\clearpage}
\begin{figure}[h]
  \begin{center}
    \hspace*{1.0cm}\includegraphics[width=0.6\linewidth]{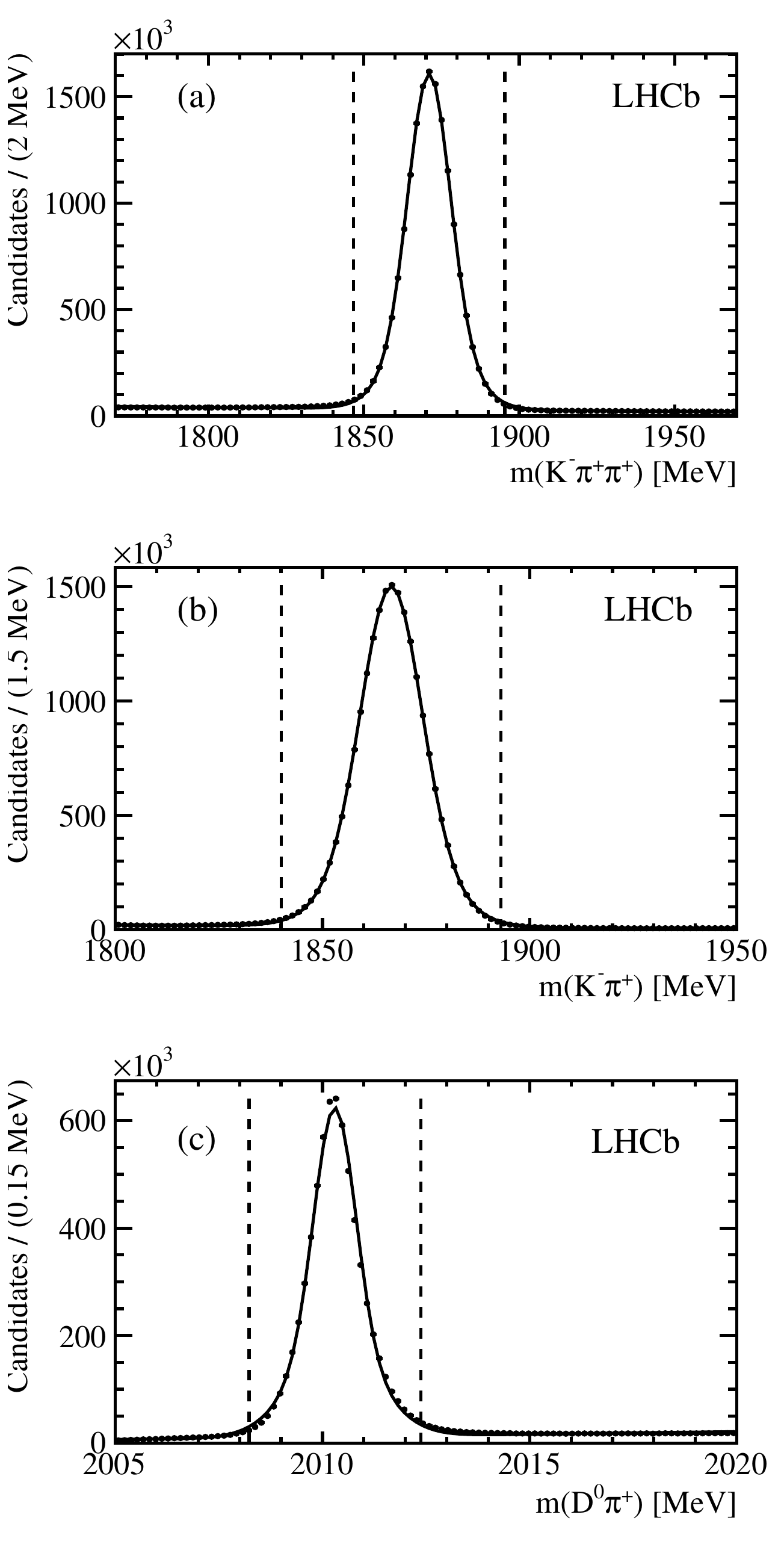}
    \vspace*{-1.0cm}
  \end{center}
  \caption{
    \small Invariant mass distribution for (a) $\Dp \to K^-\pi^+\pi^+$, (b) $\Dz \to K^-\pi^+$, and (c) $\Dstarp \to \Dz \pip$ decays. The solid lines are the results from the fits described in the text. The vertical dashed lines indicate
the signal regions.
    }
  \label{fig:fig2}
\end{figure}
The reconstructed  $D^+$, $D^0$ and $\Dstarp$ candidates are combined with all the right-sign charged pions in the event.
Each of the  $D^+ \pim$, the $D^0 \pip$, and the $\Dstarp \pim$ candidates are fitted to a common vertex with $\chi^2/{\rm ndf}<8$, where ndf is the number of degrees of freedom. 
The purity of the charmed meson candidates is enhanced by requiring the decay products to be identified by the RICH detectors, using the difference in the 
log-likelihood between the kaon and pion hypotheses $\Delta\ln\mathcal{L}_{K\pi}$~\cite{Adinolfi:2012an}. 
We require $\Delta\ln\mathcal{L}_{K\pi}>3$ for kaon tracks and a loose requirement of $\Delta\ln\mathcal{L}_{K\pi}<10$ for pions. 
In the reconstruction of $\Dp \to K^-\pi^+\pi^+$ decays, a small $\Dstarp$ signal in the $\Dz \pip$ mass spectrum is removed by demanding $\Delta m \equiv m(K^-\pi^+\pi^+) -m(K^-\pi^+) >152$~\mev.  

In order to reduce combinatorial background, the cosine of the angle between the momentum direction of the charged pion in the $D^{(*)}\pi^{\pm}$ rest frame and the momentum direction of the $D^{(*)}\pi^{\pm}$ system in the laboratory frame is required to be greater than zero. 
Due to the possible presence of multiple primary vertex candidates, 
it is required that the $D^{(*)}$ and the $\pi^{\pm}$ point to the same primary vertex.

To reduce any dependence on the mass scale, the invariant mass of the $D^{(*)}\pi^{\pm}$ system is calculated from the
measured mass difference. For example, the $\Dz \pip$ invariant mass is given by
\begin{equation}
m(\Dz \pip) = m(\Km \pip \pip) - m(\Km \pip) + m_{\Dz},
\end{equation}
where $m_{\Dz}$ is the known value of the \Dz mass~\cite{Beringer:1900zz}.  

Figure~\ref{fig:fig2} shows the $K^-\pi^+\pi^+$, $K^-\pi^+$ and $\Dz \pip$ invariant mass spectra after the selection criteria are applied.
The distributions are fitted by the sum of two Gaussian functions, with a common mean to describe the signal shape and a linear term to describe 
the combinatorial background.
The mean mass resolutions for the three distributions are 8.1, 8.8 and 0.69~\mev, respectively. 
The signal regions, indicated by the dashed vertical lines, for the \Dp, \Dz and \Dstarp candidates correspond to $\pm 3\sigma$ around the peak values and contain 15.1$\times 10^6$, 20.4$\times 10^6$ and 6.4$\times 10^6$ candidates for the $\Dp \pim$,  $\Dz \pip$ and $\Dstarp \pim$ modes, respectively.

\section{Mass spectra}
\label{sec:mass}
\begin{figure}[h]
  \begin{center}
    \hspace*{1.0cm}\includegraphics[width=0.6\linewidth]{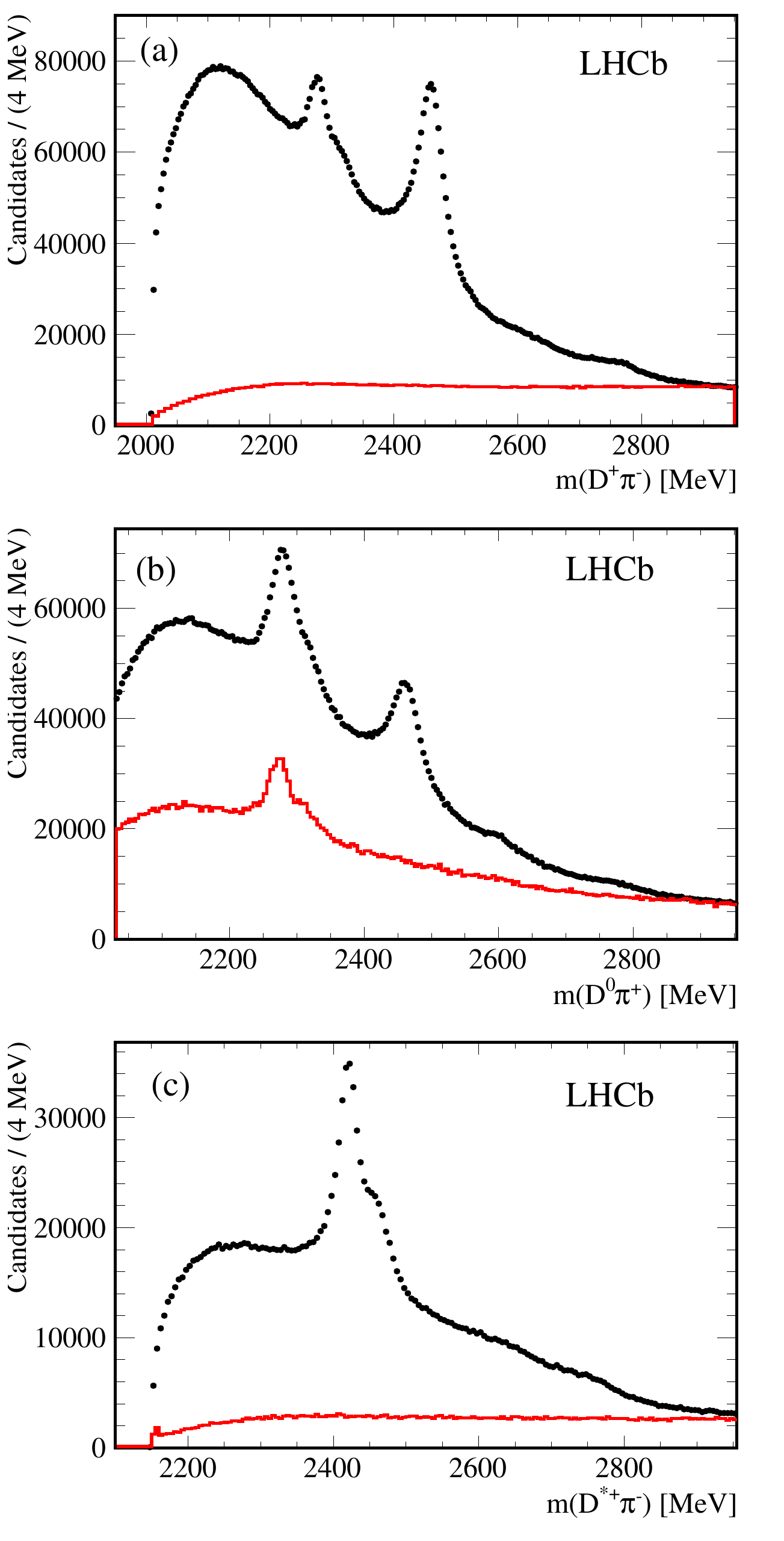}
    \vspace*{-1.0cm}
  \end{center}
  \caption{
    \small Invariant mass distribution for (a) $\Dp \pim$,  (b) $\Dz \pip$ and (c) $\Dstarp \pim$ candidates (points). The full line histograms (in red) show the wrong-sign mass spectra for (a) $\Dp \pip$,  (b) $\Dz \pim$ and (c) $\Dstarp \pip$ normalized to the same yield at high $D^{(*)}\pi$ masses.
    }
  \label{fig:fig3}
\end{figure}

The $\Dp \pim$, $\Dz \pip$ and $\Dstarp \pim$ mass spectra are shown in Fig.~\ref{fig:fig3}.
The $\Dp \pim$ and $\Dz \pip$ mass spectra evidence strong \DTwentyFourSixty signals, while in the $\Dstarp \pim$ mass spectrum clear \DTwentyFourTwentyNeutral and \DTwentyFourSixtyNeutral signals are visible.
A further reduction of the combinatorial background is achieved by performing an optimization of the signal significance and purity as a function of  
$p_{\rm T}$ of the $D^{(*)}\pi^{\pm}$ system using the well known \DTwentyFourTwenty and \DTwentyFourSixty resonances.~\footnote{We use the generic notation $D$ to indicate both neutral and charged $D$ mesons.} For this purpose, we fit the three mass spectra as explained in Sect.~7 and  Sect.~9 and obtain, for each resonance, the signal yield $N_{\rm S}$ and background yield $N_{\rm B}$ events. 
We compute the signal significance $S=N_{\rm S}/\sqrt{N_{\rm S}+N_{\rm B}}$ and signal purity $P=N_{\rm S}/(N_{\rm S}+N_{\rm B})$ and find that the requirement $p_{\rm T}(D^{(*)}\pi)>7.5\gev$
provides a good compromise between significance and purity.
After the optimization 7.9$\times 10^6$, 7.5$\times 10^6$ and 2.1$\times 10^6$ $\Dp \pim$, $\Dz \pip$ and $\Dstarp \pim$ candidates are obtained.
We also study the dependence of the signal to background ratio in the three mass spectra on the pseudorapidity of the $D^{(*)}\pi$ system and find a very weak correlation.
We analyze, for comparison and using the same selections, the wrong-sign $\Dp \pip$, $\Dz \pim$ and $\Dstarp \pip$ combinations which are also shown in Fig.~\ref{fig:fig3}.
\afterpage{\clearpage}
 
The $\Dp \pim$ mass spectrum, Fig.~\ref{fig:fig3}(a), shows a double peak structure around 2300~\mev due to cross-feed from the decay
\begin{equation}
\DTwentyFourTwentyNeutral \ {\rm or} \ \DTwentyFourSixtyNeutral \to \pi^- D^{*+} (\to D^+ \pi^0/\gamma )  \ (32.3\%),
\label{dp_pim}
\end{equation}
where the $\pi^0/\gamma$ is not reconstructed; the last number, in parentheses, indicates the branching fraction of $D^{*+} \to  D^+ \pi^0/\gamma$ decays~\cite{Beringer:1900zz}.
We observe a strong \DTwentyFourSixtyNeutral signal and weak structures around 2600 and 2750~\mev. The wrong-sign $\Dp \pip$ mass spectrum does not show any structure.

The $\Dz \pip$ mass spectrum, Fig.~\ref{fig:fig3}(b), shows an enhanced double peak structure around 2300~\mev
due to cross-feed from the decays
\begin{equation}
\DTwentyFourTwentyCharged  \ {\rm or} \ \DTwentyFourSixtyCharged \begin{array}{l} \to \pi^+ D^{*0} \begin{array}{l}  \\ ( \to  D^0 \pi^0)  \ (61.9\%) \\ (\to  D^0 \gamma)  \ (38.1\%) \ . \end{array} \end{array} 
\label{dz_pip}
\end{equation}
The \DTwentyFourSixtyCharged signal and weak structures around 2600 and 2750~\mev are observed. 
In comparison, the wrong-sign $\Dz \pim$ mass spectrum does 
show the presence of structures in the 2300~\mev mass region, similar to those observed in the $\Dz \pip$ mass spectrum.
These structures are due to cross-feed from the decay
\begin{equation}
\DTwentyFourTwentyNeutral \ {\rm or} \ \DTwentyFourSixtyNeutral \to \pi^- D^{*+} ( \to   D^0 \pi^+)  \ (67.7\%) \ .
\label{dz_pim}
\end{equation}

The $\Dstarp \pim$ mass spectrum, Fig.~\ref{fig:fig3}(c), is dominated by the presence of the \DTwentyFourTwentyNeutral and \DTwentyFourSixtyNeutral signals. At higher mass, complex broad structures are evident in the mass region between 2500 and 2800~\mev.

\section{Simulation}
\label{sec:mc}

Simulated events are used to study the effects of the detector on the observed mass resolution. The $pp$ collisions are generated using
\pythia~6.4~\cite{Sjostrand:2006za} with a specific \lhcb
configuration~\cite{LHCb-PROC-2010-056}. Decays of hadronic particles
are described by \evtgen~\cite{Lange:2001uf} and the
interaction of the generated particles with the detector and its
response are implemented using the \geant
toolkit~\cite{Allison:2006ve,Agostinelli:2002hh} as described in
Ref.~\cite{LHCb-PROC-2011-006}. 

Simulated events are reconstructed in the same manner as data.
We analyze samples of full detector simulations to estimate the reconstruction efficiency, mass resolution and possible bias in the reconstruction chain. 
We also make use of simple event generator level simulations~\cite{genbod} to study kinematic effects. The tight trigger conditions and selection criteria have the effect of producing very
low integrated efficiencies, which we calculate to be (0.149$\pm$0.004)\%, (0.056$\pm$0.005)\% and (0.064$\pm$0.003)\% for $\Dp \pim$, $\Dz \pip$ and $\Dstarp \pim$ candidates, respectively.

To estimate the detector resolution we compare generated and reconstructed invariant masses and obtain experimental resolutions as functions of the reconstructed mass. The analysis of these simulated samples shows no bias in the reconstructed invariant masses. We estimate resolutions which, in the mass region between 2000 and 2900~\mev, are similar for the three mass spectra and 
range from 1.0 to 4.5~\mev as a function of the mass. Since the widths of the resonances appearing in the three mass spectra are much larger than the experimental resolutions, resolution effects are neglected.

\section{Mass fit model}
\label{sec:fit}  

Binned $\chi^2$ fits to the three mass spectra are performed.
The \DTwentyFourSixty and \DTwentyFourHundred signal shapes in two-body decays are parameterized with a relativistic Breit-Wigner that includes the mass-dependent factors for a D-wave and S-wave decay, respectively.
The radius entering in the Blatt-Weisskopf~\cite{BW} form factor is fixed to 4~$\gev^{-1}$. 
Other resonances appearing in the mass spectra are described by Breit-Wigner lineshapes. 
All Breit-Wigner expressions are multiplied by two-body phase space.
The cross-feed lineshapes from \DTwentyFourTwenty and \DTwentyFourSixty appearing in the $\Dp \pim$ and $\Dz \pip$ mass spectra are described by a Breit-Wigner function fitted to the data with the parameters given in Table~\ref{tab:feed}.
Resonances are included sequentially in order to test the $\chi^2$ improvement when a new contribution is included.
The background $B(m)$ is described by an empirical shape~\cite{delAmoSanchez:2010vq} 
{\begin{eqnarray}
B(m) = & P(m)e^{a_1m+a_2m^2} \ {\rm for} \ m<m_0, \nonumber\\
B(m) = & P(m)e^{b_0+b_1m+b_2m^2} \ {\rm for} \ m>m_0,
\end{eqnarray}
where $P(m)$ is the two-body phase space and $m_0$ is a free parameter.
\begin{table}[t]
\caption{\small Breit-Wigner parameters describing the cross-feed from \DTwentyFourTwenty and \DTwentyFourSixty in the $\Dp \pim$ and $\Dz \pip$ final states.}
\label{tab:feed}
\begin{center}
\begin{tabular}{llrr}
Final state & Parameter (\mev) & \DTwentyFourTwenty & \DTwentyFourSixty \cr
\hline
$\Dp \pim$ &  Mass  & 2276.5 & 2319.8  \cr
 & Width  & 38.3 & 50.0 \cr
\hline
$\Dz \pip$ &  Mass & 2278.4 & 2319.4  \cr
 & Width  & 44.9 & 49.1 \cr
\hline
\end{tabular}
\end{center}
\end{table}
\begin{table}[b]
\caption{\small Definition of the categories selected by different ranges of \cthetah, and fraction of the total natural parity contribution.}
\label{tab:nat_unnat}
\begin{center}
\begin{tabular}{llc}
Category & Selection & natural parity fraction (\%)\cr
\hline
{\it Enhanced unnatural parity sample} & $|\cos \theta_{\rm H}| > 0.75 $ & \ 8.6 \cr
{\it Natural parity sample} &  $|\cos \theta_{\rm H}|<0.5$ & 68.8 \cr
{\it Unnatural parity sample} & $|\cos \theta_{\rm H}|>0.5$ & 31.2 \cr        
\hline    
\end{tabular}
\end{center}
\end{table}

The two functions and their first derivatives are required to be continuous at $m_0$ such that
\begin{equation}
b_1 = a_1 + 2 \ m_0(a_2 - b_2) \ ,
\end{equation}
\begin{equation}
b_0 = m_0(a_1 - b_1) + m_0^2(a_2 - b_2) \ .
\end{equation}
Therefore the background model has four free parameters: $m_0, \ a_1, \ a_2$ and $b_2$.

\vspace{0.5cm}
\begin{boldmath}
\section{Fit to the $D^{*+}\pi^-$ mass spectrum}
\end{boldmath}

\label{sec:fit_dstarpi}

\begin{figure}[hb]
  \begin{center}
    \hspace*{1.0cm}\includegraphics[width=0.80\linewidth]{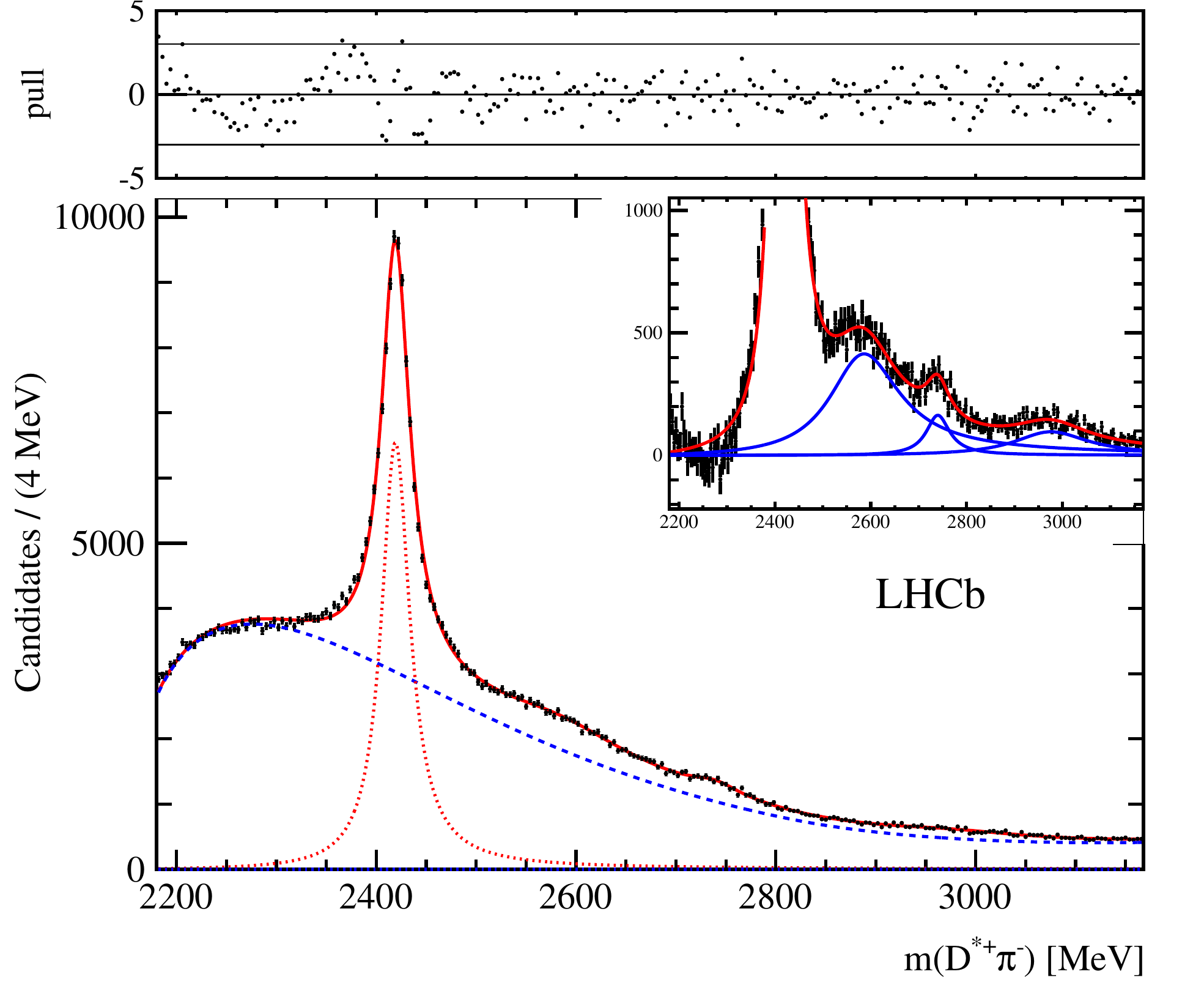}
    \vspace*{-0.5cm}
  \end{center}
  \caption{
    \small Fit to the $\Dstarp \pim$ mass spectrum, {\it enhanced unnatural parity sample}, as defined in Table~\ref{tab:nat_unnat}.  The dashed (blue) line shows the fitted background, the dotted (red) line
the \DTwentyFourTwentyNeutral contribution. The inset displays the $\Dstarp \pim$ mass spectrum after subtracting the fitted background. The full line curves (blue) show the contributions from \DTwentyFiveFiftyNeutral, \DTwentySevenFiftyNeutral, and \DThreeU states. The top window displays the pull distribution where the horizontal lines indicate $\pm 3 \sigma$. The pull is defined as $(N_{\rm data} - N_{\rm fit})/\sqrt{N_{\rm data}}$.
    }
  \label{fig:fig4}
\end{figure}

\begin{figure}[ht]
  \begin{center}
    \hspace*{1.0cm}\includegraphics[width=0.80\linewidth]{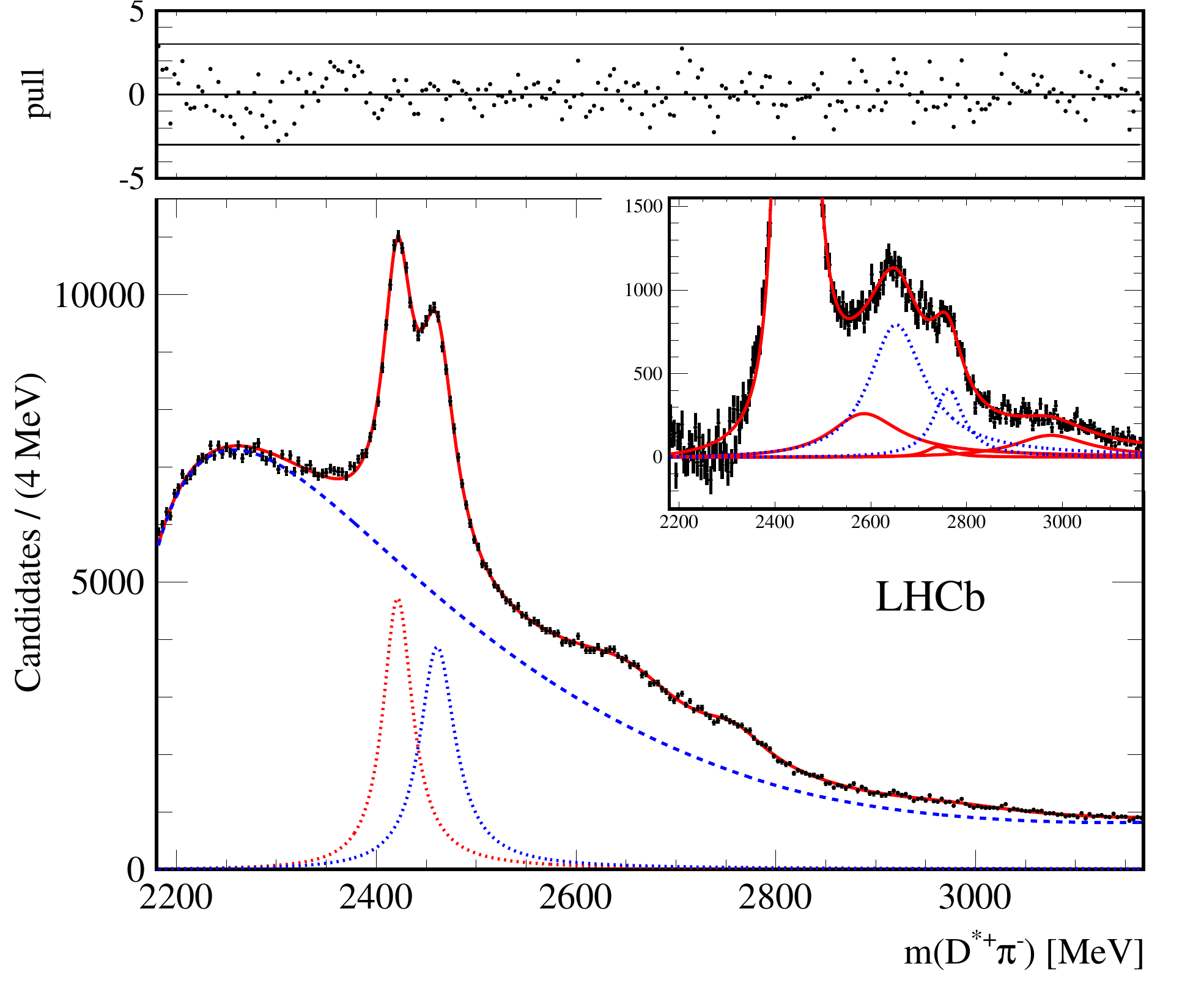}
    \vspace*{-0.5cm}
  \end{center}
  \caption{
    \small Fit to the {\it natural parity sample} $\Dstarp \pim$ mass spectrum. The dashed (blue) line shows the fitted background, the dotted lines 
the \DTwentyFourTwentyNeutral (red) and  \DTwentyFourSixtyNeutral (blue) contributions. The inset displays the $\Dstarp \pim$ mass spectrum after subtracting the fitted background. The full line curves (red) show the contributions from \DTwentyFiveFiftyNeutral, \DTwentySevenFiftyNeutral, and \DThreeU. The dotted (blue) lines display the \DTwentySixHundredNeutral and \DTwentySevenSixtyNeutral contributions. The top window shows the pull distribution where the horizontal lines indicate $\pm 3 \sigma$.
    }
  \label{fig:fig5}
\end{figure}

\begin{figure}[ht]
  \begin{center}
    \hspace*{1.0cm}\includegraphics[width=0.8\linewidth]{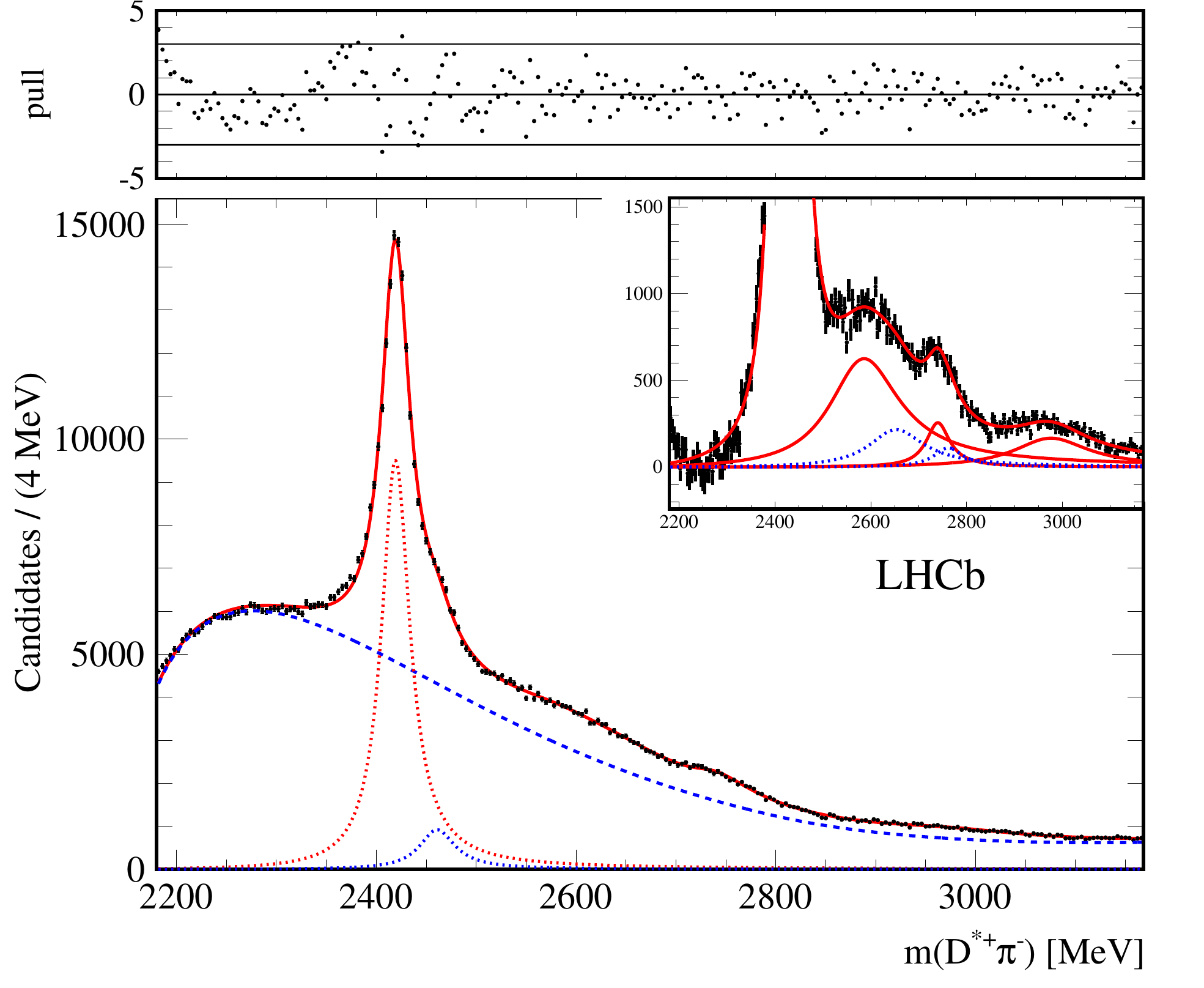}
    \vspace*{-0.5cm}
  \end{center}
  \caption{
    \small Fit to the {\it unnatural parity sample} $\Dstarp \pim$ mass spectrum.  The dashed (blue) line shows the fitted background, the dotted lines
the \DTwentyFourTwentyNeutral (red) and  \DTwentyFourSixtyNeutral (blue) contributions. The inset displays the $\Dstarp \pim$ mass spectrum after subtracting the fitted background. The full line curves (red) show the contributions from \DTwentyFiveFiftyNeutral, \DTwentySevenFiftyNeutral, and \DThreeU states. The dotted (blue) lines display the \DTwentySixHundredNeutral and \DTwentySevenSixtyNeutral contributions. The top window shows the pull distribution where the horizontal lines indicate $\pm 3 \sigma$.
    }
  \label{fig:fig6}
\end{figure}

\begin{figure}[ht]
  \begin{center}
    \hspace*{1.0cm}\includegraphics[width=0.8\linewidth]{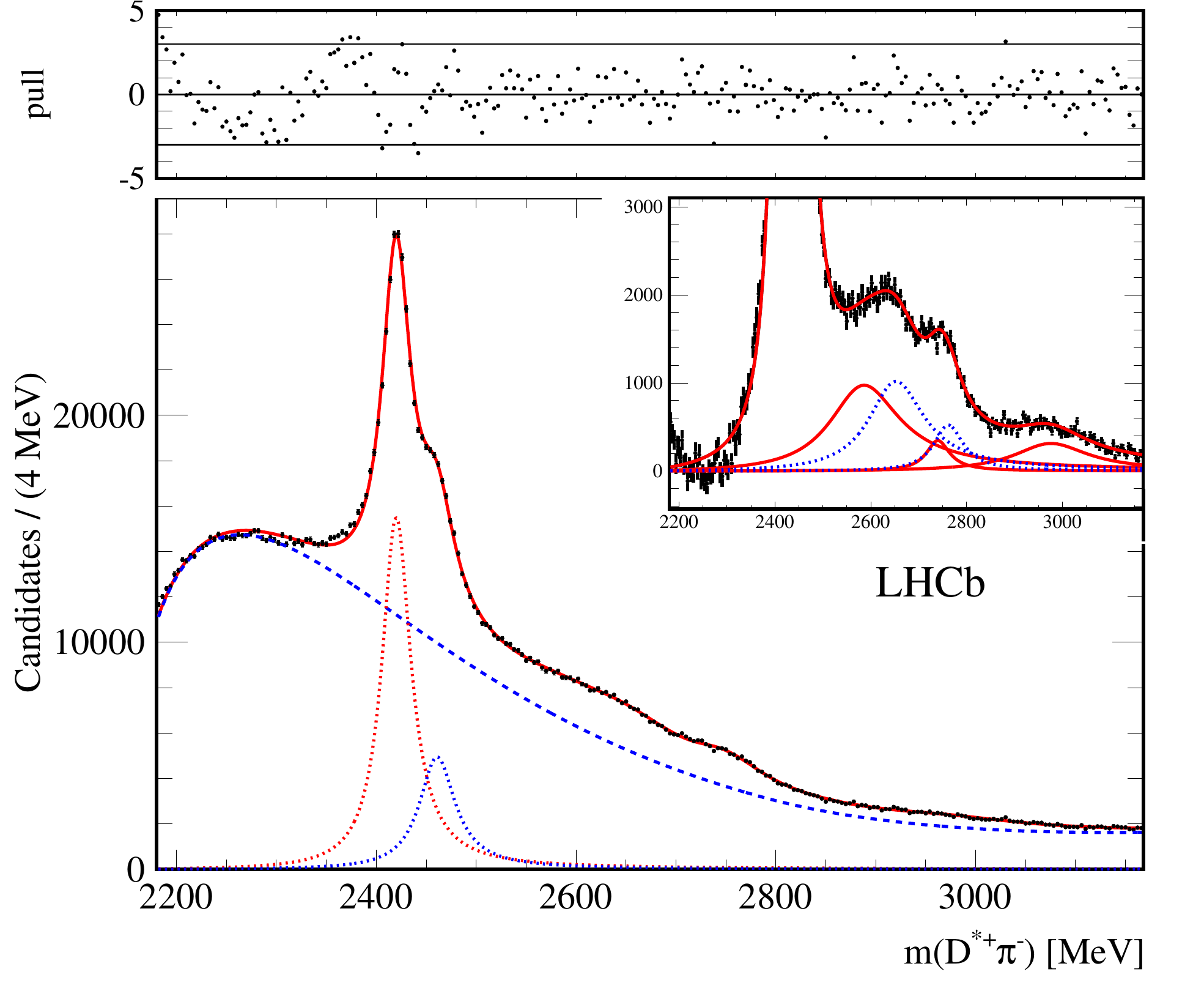}
    \vspace*{-0.5cm}
  \end{center}
  \caption{
    \small Fit to the total $\Dstarp \pim$ sample.  The dashed (blue) line shows the fitted background, the dotted lines 
the \DTwentyFourTwentyNeutral (red) and  \DTwentyFourSixtyNeutral (blue) contributions. The inset displays the $\Dstarp \pim$ mass spectrum after subtracting the fitted background. The full line curves (red) show the contributions from \DTwentyFiveFiftyNeutral, \DTwentySevenFiftyNeutral, and \DThreeU states. The dotted (blue) lines display the \DTwentySixHundredNeutral and \DTwentySevenSixtyNeutral contributions. The top window shows the pull distribution where the horizontal lines indicate $\pm 3 \sigma$.
    }
  \label{fig:fig7}
\end{figure} 
\begin{table}[t]
\caption{\small Mass intervals, number of bins, yields,  and $\chi^2/{\rm ndf}$ in the fits to the different mass spectra.}
\label{tab:yields}
\begin{center}
\begin{tabular}{llcccc}
Final state & Selection & Fit Range & Number & Candidates & $\chi^2/{\rm ndf}$ \cr
            &           &   (MeV)            & of bins& ($\times 10^6$)      & \cr
\hline
$\Dp \pim$ & Total & 2050-3170& 280 & 7.90 & 551/261 \cr
\hline
$\Dz \pip$ &Total & 2050-3170& 280 & 7.50 & 351/262 \cr
\hline
$\Dstarp \pim$ &Total &  2180-3170 &247  & 2.04 & 438/234\cr
$\Dstarp \pim$ & {\it Natural} & & & 0.98 & 263/229 \cr
             & {\it parity sample} & & & & \cr
$\Dstarp \pim$ &{\it Unnatural}  & & & 1.06 & 364/234\cr
             & {\it parity sample} & & & & \cr
$\Dstarp \pim$ & {\it Enhanced unnatural parity} & & &  0.55 & 317/230\cr
             & {\it sample} & & & & \cr
\hline
\end{tabular}
\end{center}
\end{table}

Due to the three-body decay and the availability of the helicity angle information, the fit to the $\Dstarp \pim$ mass spectrum allows a spin analysis of the produced resonances and a 
separation of the different spin-parity components. We define the helicity angle \mthetah as the angle between the $\pim$ and the $\pip$ from the $\Dstarp$ decay, in the rest frame of the $\Dstarp \pim$ system. 
Full detector simulations are used to measure the efficiency as a function of \mthetah, which is found to be uniform. 

It is expected that the angular distributions are proportional to $\sin^2\mthetah$ for natural parity resonances and proportional to $1+h\cos^2\mthetah$ for unnatural parity resonances, where $h>0$ is a free parameter. 
The $\Dstar \pi$ decay of a $J^P=0^+$ resonance is forbidden. 
Therefore candidates
selected in different ranges of \cthetah can enhance or suppress the different spin-parity contributions. We separate the $\Dstarp \pim$ data into three different categories, summarized in Table~\ref{tab:nat_unnat}. The candidate yields for these categories are given in Table~\ref{tab:yields}, which also reports the mass intervals, the number of bins, and the resulting $\chi^2/{\rm ndf}$ in the fits to the different mass spectra.

The data and fit for the $\Dstarp \pim$  {\it enhanced unnatural parity sample} are shown in Fig.~\ref{fig:fig4} and the resulting fit
parameters are summarized in Table~\ref{tab:fits}. 
The mass spectrum is dominated by the presence of the unnatural parity \DTwentyFourTwentyNeutral resonance. 
The fitted natural parity \DTwentyFourSixtyNeutral contribution is consistent with zero, as expected. 
To obtain a good fit to the mass spectrum, three further resonances are needed. We label them \DTwentyFiveFiftyNeutral, \DTwentySevenFiftyNeutral, and \DThreeU.
The presence of these states in this sample indicates unnatural parity assignments.

The masses and widths of the unnatural parity resonances are fixed in the fit to the {\it natural parity sample}.
The fit is shown in Fig.~\ref{fig:fig5} and the obtained resonance 
parameters are summarized in Table~\ref{tab:fits}. 
The mass spectrum shows that the unnatural parity resonance \DTwentyFourTwentyNeutral is suppressed with respect to that observed in the
 {\it enhanced unnatural parity sample}. 
There is a strong contribution of the natural parity \DTwentyFourSixtyNeutral resonance and
contributions from the \DTwentyFiveFiftyNeutral, \DTwentySevenFiftyNeutral and \DThreeU states.
To obtain a good fit, two additional resonances are needed, which we label \DTwentySixHundredNeutral and \DTwentySevenSixtyNeutral.

The {\it unnatural parity sample} is used as a cross-check. In this fit, the parameters of all the resonances are fixed to the values 
obtained from the previous fits.  The fit is shown in Fig.~\ref{fig:fig6}. We observe, as expected, small contributions from the natural parity resonances. We also fit the total $\Dstarp \pim$ mass spectrum, again with all the resonance parameters fixed. The data and fit are shown in Fig.~\ref{fig:fig7}.

Table~\ref{tab:fits} summarizes the measured resonance parameters and yields. The resonance parameters are obtained from the fits to the {\it enhanced unnatural parity sample} and {\it natural parity sample}, apart for the parameters of the \DTwentyFourTwentyNeutral resonance, which are extracted from the fit to the total sample.
The significances are computed as $\sqrt{\Delta \chi^2}$ where $\Delta \chi^2$ is 
the difference between the $\chi^2$ values when a resonance is included or excluded from the fit while all the other resonances parameters are allowed to vary.  All the statistical significances are well above 5$\sigma$.

\begin{sidewaystable}
\caption{\small Resonance parameters, yields and statistical significances. The first uncertainty is statistical, the second systematic.
Significances are evaluated using the method described in the text.}
\label{tab:fits}
\begin{center}
\begin{tabular}{c | c | r@{}c@{}l | r@{}c@{}l | r@{}c@{}l | c}
Resonance & Final state & \multicolumn{3}{c|}{Mass (MeV)} & \multicolumn{3}{c|}{Width (MeV)} & \multicolumn{3}{c|}{Yields $\times 10^3$} & Significance ($\sigma$)  \cr
\hline
\DTwentyFourTwentyNeutral & $\Dstarp \pim$ & 2419.6 $\pm$ & \, 0.1 \, &  $\pm$ 0.7 &  35.2 $\pm$ & \, 0.4 \, &  $\pm$ 0.9  &  210.2 $\pm$ & \, 1.9 \, &  $\pm$ 0.7 &   \,   \\
\DTwentyFourSixtyNeutral  & $\Dstarp \pim$ & 2460.4 $\pm$ & \, 0.4 \, &  $\pm$ 1.2 &  43.2 $\pm$ & \, 1.2 \, &  $\pm$ 3.0  &   81.9 $\pm$ & \, 1.2 \, &  $\pm$ 0.9 &   \,    \\
\DTwentySixHundredNeutral & $\Dstarp \pim$ & 2649.2 $\pm$ & \, 3.5 \, &  $\pm$ 3.5 & 140.2 $\pm$ & \, 17.1 \,&  $\pm$ 18.6 &   50.7 $\pm$ & \, 2.2 \, &  $\pm$ 2.3 &  24.5     \\
\DTwentySevenSixtyNeutral & $\Dstarp \pim$ & 2761.1 $\pm$ & \, 5.1 \, &  $\pm$ 6.5 &  74.4 $\pm$ & \, 3.4 \, &  $\pm$ 37.0 &   14.4 $\pm$ & \, 1.7 \, &  $\pm$ 1.7 &  10.2     \\
\DTwentyFiveFiftyNeutral  & $\Dstarp \pim$ & 2579.5 $\pm$ & \, 3.4 \, &  $\pm$ 5.5 & 177.5 $\pm$ & \, 17.8 \, & $\pm$ 46.0 &   60.3 $\pm$ & \, 3.1 \, &  $\pm$ 3.4 &  18.8   \\
\DTwentySevenFiftyNeutral & $\Dstarp \pim$ & 2737.0 $\pm$ & \, 3.5 \, &  $\pm$11.2 &  73.2 $\pm$ & \, 13.4 \, & $\pm$ 25.0 &    7.7 $\pm$ & \, 1.1 \, &  $\pm$ 1.2 &  \ 7.2     \\
\DThreeU                  & $\Dstarp \pim$ & 2971.8 $\pm$ & \, 8.7 \, &            & 188.1 $\pm$ & \, 44.8 \, &            &    9.5 $\pm$ & \, 1.1 \, &             &  \ 9.0     \\
\hline
\DTwentyFourSixtyNeutral  & $\Dp \pim$     & 2460.4 $\pm$ & \, 0.1 \, & $\pm$ 0.1  &  45.6 $\pm$ & \,  0.4 \, & $\pm$ 1.1 &   675.0 $\pm$ & \, 9.0 \, &  $\pm$ 1.3 &                   \\
\DTwentySevenSixtyNeutral & $\Dp \pim$     & 2760.1 $\pm$ & \, 1.1 \, & $\pm$ 3.7  &  74.4 $\pm$ & \,  3.4 \, & $\pm$19.1 &    55.8 $\pm$ & \, 1.3 \, &  $\pm$ 10.0 &   17.3     \\
\DThreeNeutral            & $\Dp \pim$     & 3008.1 $\pm$ & \, 4.0 \, &            & 110.5 $\pm$ & \, 11.5 \, &            &   17.6 $\pm$  & \,1.1 \, &             &   21.2   \\
\hline 
\DTwentyFourSixtyCharged  & $\Dz \pip$     & 2463.1 $\pm$ & \, 0.2 \, & $\pm$ 0.6  &  48.6 $\pm$ & \,  1.3 \, & $\pm$ 1.9 &   341.6 $\pm$ & \,22.0 \, &  $\pm$ 2.0 &                   \\
\DTwentySevenSixtyCharged & $\Dz \pip$     & 2771.7 $\pm$ & \, 1.7 \, & $\pm$ 3.8  &  66.7 $\pm$ & \,  6.6 \, & $\pm$10.5 &    20.1 $\pm$ & \, 2.2 \, &  $\pm$ 1.0 &   18.8      \\
\DThreeCharged            & $\Dz \pip$     & 3008.1       & \, (fixed) \,   &        & 110.5       & \, (fixed) \, &           &    7.6 $\pm$  & \, 1.2 \, &             & \  6.6    \\
\hline
\end{tabular}
\end{center}
\end{sidewaystable}

\vspace{0.5cm}
\begin{boldmath}
\section{Spin-parity analysis of the \DstarPi system}
\end{boldmath}
\label{sec:spin}

In order to obtain information on the spin-parity assignment of the states observed in the  $\Dstarp \pim$ mass spectrum,
the data are subdivided into ten equally spaced bins in $\cthetah$. The ten mass spectra are then fitted
with the model described above with fixed resonance parameters to obtain the yields as functions of $\cthetah$ for each resonance.

The resulting distributions for \DTwentyFourTwentyNeutral and \DTwentyFourSixtyNeutral are shown in Fig.~\ref{fig:fig8}. They have been fitted using the functions
described in Table~\ref{tab:spin}. A good description of the data is obtained in terms of the expected angular distributions for $J^P=1^+$ and $J^P=2^+$ resonances.

\begin{figure}[ht]
  \begin{center}
    \hspace*{0.5cm}\includegraphics[width=0.8\linewidth]{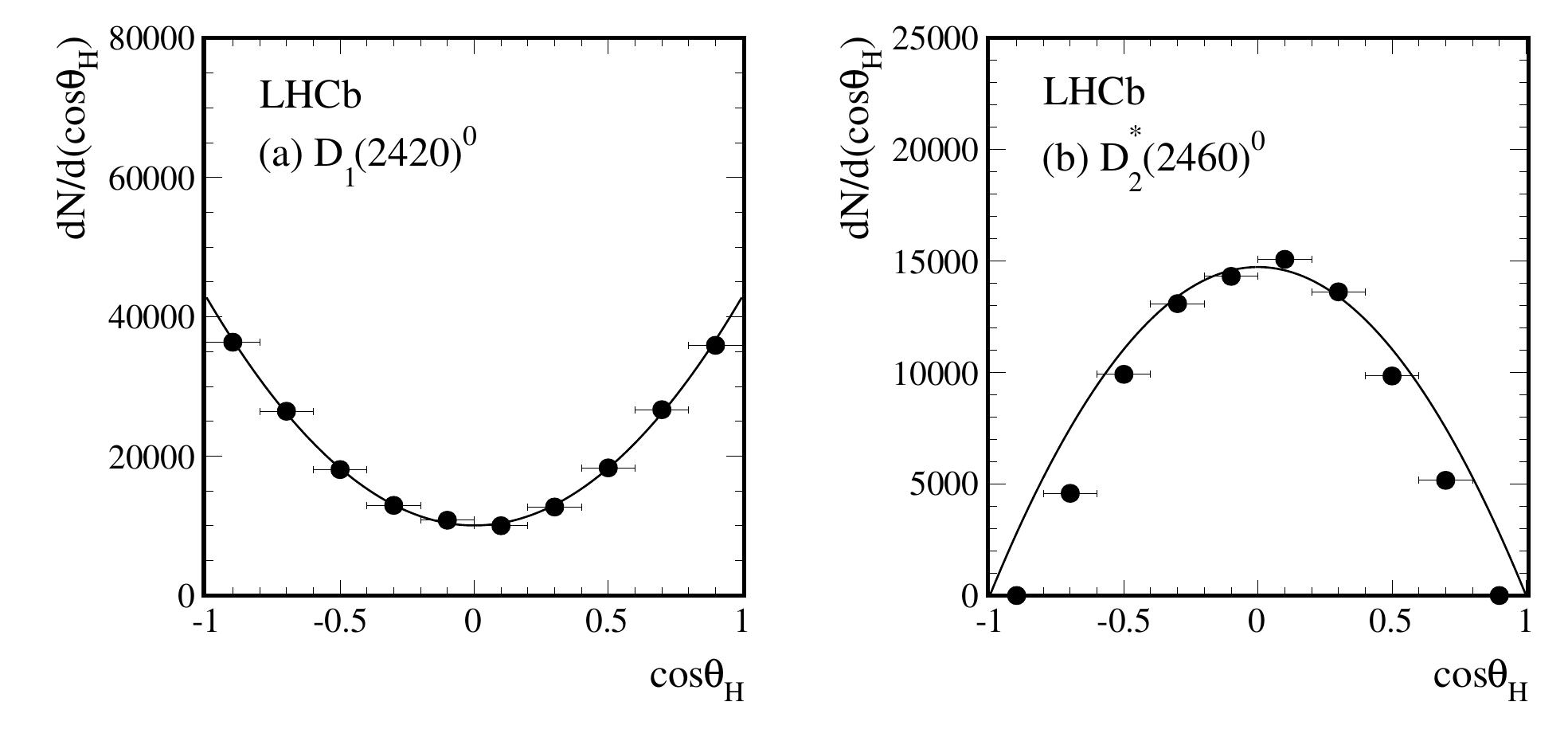}
    \vspace*{-0.5cm}
  \end{center}
  \caption{
    \small Distributions of (a) \DTwentyFourTwentyNeutral and (b) \DTwentyFourSixtyNeutral candidates as functions of the 
helicity angle \cthetah. The distributions are fitted with unnatural and natural parity functions, respectively.
    }
  \label{fig:fig8}
\end{figure}

Figure~\ref{fig:fig9} shows the resulting distributions for the \DTwentySixHundredNeutral and \DTwentySevenSixtyNeutral states. In this case we compare the distributions with expectations from natural parity, unnatural parity and $J^P=0^-$. In the case of unnatural parity, the $h$ parameter, in $ 1+h\cos^2\theta_{\rm H}$,  is constrained to be positive and therefore the fit gives $h=0$. In both cases, the distributions are best fitted by the 
natural parity hypothesis.

Figure~\ref{fig:fig10} shows the angular distributions for the \DTwentyFiveFiftyNeutral, \DTwentySevenFiftyNeutral and \DThreeU states. The distributions are fitted with 
natural parity and unnatural parity. The $J^P=0^-$ hypothesis is also considered for \DTwentyFiveFiftyNeutral. The results from the fits are given in Table~\ref{tab:spin}. In all cases unnatural parity is preferred over
a natural parity assignment.
\begin{figure}[ht]
  \begin{center}
    \hspace*{0.5cm}\includegraphics[width=0.8\linewidth]{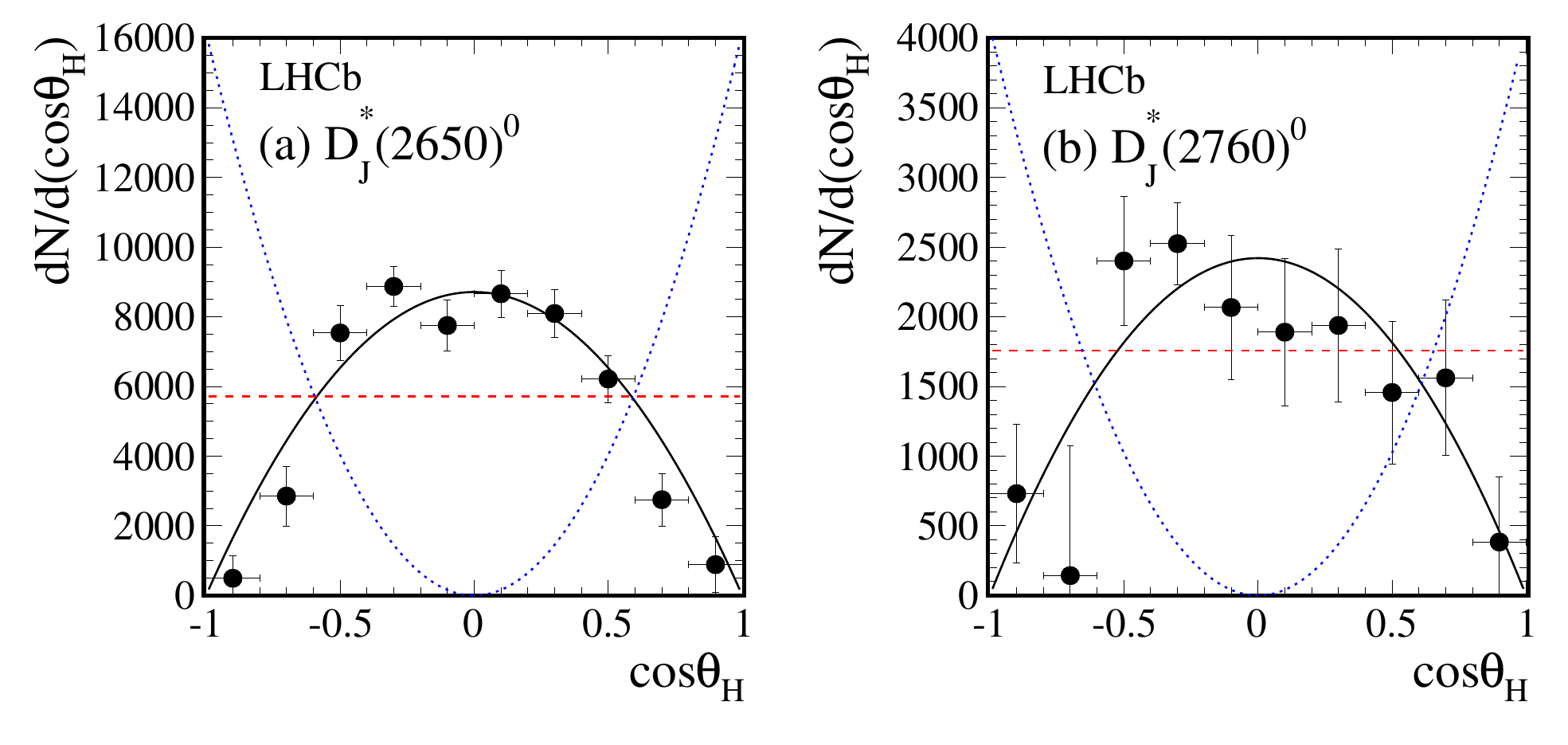}
    \vspace*{-0.5cm}
  \end{center}
  \caption{
    \small  Distributions of (a) \DTwentySixHundredNeutral and (b) \DTwentySevenSixtyNeutral candidates as functions of the 
helicity angle \cthetah. The distributions are fitted with 
natural parity (black continuous), unnatural parity (red, dashed) and $J^P=0^-$ (blue, dotted) functions.
    }
  \label{fig:fig9}
\end{figure}

\begin{figure}[ht]
  \begin{center}
    \hspace*{0.1cm}\includegraphics[width=1.0\linewidth]{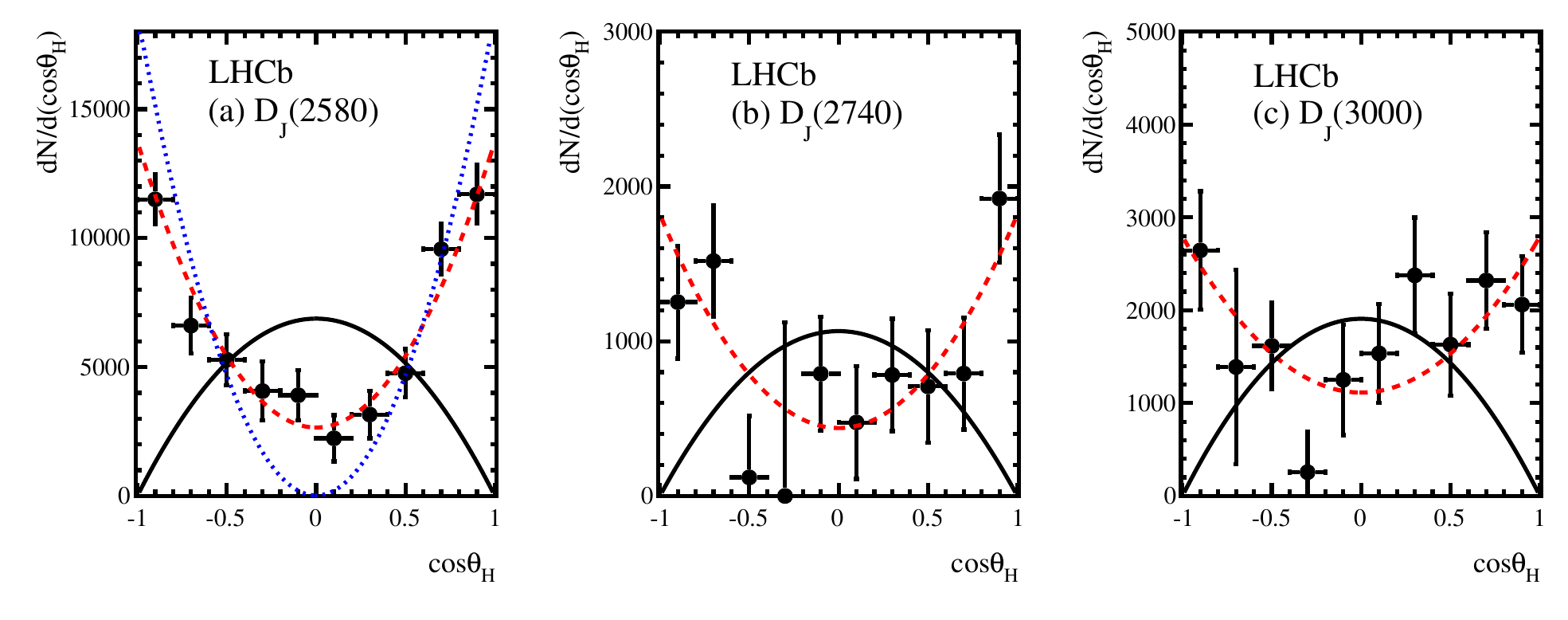}
    \vspace*{-0.5cm}
  \end{center}
  \caption{
    \small Distributions of (a) \DTwentyFiveFiftyNeutral, (b) \DTwentySevenFiftyNeutral and (c) \DThreeU candidates as functions of the 
helicity angle \cthetah. The distributions are fitted with 
natural parity (black continuous) and unnatural parity (red, dashed) functions. In (a) the $J^P=0^-$ (blue, dotted) hypothesis is also tested. 
    }
  \label{fig:fig10}
\end{figure}
\begin{sidewaystable}
\caption{\small Results from the fits to the $\cthetah$ angular distributions for the resonances observed in the $\Dstarp \pim$ mass spectrum. The resulting $\chi^2/{\rm ndf}$ for different spin-parity assignments are reported. For each resonance, the expected angular distributions are indicated, where $h$ is a free parameter.
The favoured spin-parity assignment is indicated in bold font.}
\label{tab:spin}
\begin{center}
\begin{tabular}{c|c|c|c|c|c|c|c}
Resonance  & $J^P$ & $\chi^2/{\rm ndf}$ & $J^P$ & $\chi^2/{\rm ndf}$ & $J^P$ & $\chi^2/{\rm ndf}$ & $h$ Parameter \cr
           & Function & & Function &  & Function & & \cr
\hline
\DTwentyFourTwentyNeutral & $1^{+}$ &  0.67/8 & & & &  &3.30 $\pm$ 0.48    \cr 
 & $ 1+h\cos^2\theta_{\rm H}$ & & & & & & \cr
\hline
\DTwentyFourSixtyNeutral & $2^{+}$ & 8.5/9 & & & & &  \cr
 & $ \sin^2\theta_{\rm H}$ & & & & & & \cr
\hline
\DTwentySixHundredNeutral & {\bf Natural} &  6.8/9 & unnatural & 200/9 & $0^-$ & 342/9 & \cr
 & $ \sin^2\theta_{\rm H}$ &  & Const. & &  $ \cos^2\theta_{\rm H}$ & & \cr
\hline
\DTwentySevenSixtyNeutral & {\bf Natural} & 5.8/9  & unnatural & 26/9 & $0^-$ & 94/9 & \cr
 & $ \sin^2\theta_{\rm H}$ &  & Const. & &  $ \cos^2\theta_{\rm H}$ & & \cr
\hline
\DTwentyFiveFiftyNeutral & natural & 151/9  & {\bf Unnatural} & 3.4/8  & $0^-$ & 23/9 & 4.2 $\pm$ 1.3\cr
 & $ \sin^2\theta_{\rm H}$ &  & $ 1+h\cos^2\theta_{\rm H}$ & &  $ \cos^2\theta_{\rm H}$ & & \cr
\hline
\DTwentySevenFiftyNeutral & natural & 34/9 & {\bf Unnatural} &  6.6/8 &  & & 3.1 $\pm$ 2.2\cr 
 & $ \sin^2\theta_{\rm H}$ &  & $ 1+h\cos^2\theta_{\rm H}$ & &  & & \cr
\hline
\DThreeU & natural &  36.6/9 & {\bf Unnatural} & 10/8 & &  & 1.5 $\pm$ 0.9\cr
 & $ \sin^2\theta_{\rm H}$ &  & $ 1+h\cos^2\theta_{\rm H}$ & &  & & \cr
\hline
\end{tabular}
\end{center}
\end{sidewaystable}

\vspace{0.5cm}
\begin{boldmath}
\section{Fit to the \DPi and \DzPi mass spectra}
\end{boldmath}
\label{sec:fit_dpi}

The $\Dp \pim$ and $\Dz \pip$ mass spectra consist of natural parity resonances.
However these final states are affected by cross-feed from all the resonances that decay to the $\Dstar \pi$ final state. Figures~\ref{fig:fig3}(a) and (b) show 
(in the mass region around 2300 MeV) cross-feed contributions from 
\DTwentyFourTwenty and \DTwentyFourSixty decays. However we also expect (in the mass region between 2400 and 2600 MeV) the presence of structures originating from the complex resonance structure present in the $\Dstar \pi$ mass spectrum in the mass region between 2500 and 2800~\mev.

To obtain an estimate of the lineshape and size of the cross-feed, we normalize the $\Dstarp \pim$ mass spectrum to the $\Dp \pim$ mass spectrum using 
the sum of the \DTwentyFourTwentyNeutral and \DTwentyFourSixtyNeutral yields in the $\Dstarp \pim$ mass spectrum ($N_{\rm sig}$) and the sum of the cross-feed in the 
$\Dp \pim$ mass spectrum ($N^{\rm feed}_{\Dp \pim}$). We estimate that each resonance appearing in the $\Dstarp \pim$ should also appear in the $\Dp \pim$ mass spectrum with a yield given by
\begin{equation}
 N(\Dp \pim) = N(\Dstarp \pim) R_{\Dp \pim} \ ,
\end{equation}
where $R_{\Dp \pim}=N^{\rm feed}_{\Dp \pim}/N_{\rm sig}$.
Here $N(\Dstarp \pim)$ is the yield measured in the $\Dstarp \pim$ final state, $N(\Dp \pim)$ is the expected yield in the $\Dp \pim$ mass spectrum and $R_{\Dp \pim}=1.41 \pm 0.02$ where the uncertainty is statistical only. 

Assuming similar yields for the \DTwentyFourTwentyCharged and \DTwentyFourSixtyCharged resonances, we estimate for the $\Dz \pip$ channel,
\begin{equation}
 N(\Dz \pip) = N(\Dstarp \pim) R_{\Dz \pip} \ ,
\end{equation}
where $R_{\Dz \pi^+}=N^{\rm feed}_{\Dz \pip}/N_{\rm sig} =1.87 \pm 0.02$ is the corresponding value for the $\Dz \pip$ channel.

To obtain the expected lineshape of the cross-feed in the $\Dp \pim$ final state, we perform a study based on a generator level simulation.
We generate \DTwentySixHundredNeutral, \DTwentySevenSixtyNeutral, \DTwentyFiveFiftyNeutral and \DTwentySevenFiftyNeutral decays according to the chain described in Eq.~(\ref{dp_pim}).
Given the small branching fraction of the $\Dstarp \to \Dp \gamma$ decay, (1.6 $\pm$ 0.4)\%, we only generate the $\Dstarp \to \Dp \piz$ decay.
The parameters of the resonances are as reported in Table~\ref{tab:fits} and the decays to $\Dstarp \pim$ are uniform over phase space. We then compute the resulting $\Dp \pim$ mass spectra and normalize each contribution to the measured yields. The overall resulting structures
are then scaled by the factor $R_{\Dp \pim}$ and superimposed on the $\Dp \pim$ mass spectrum shown in Fig.~\ref{fig:fig11}.

\begin{figure}[t]
  \begin{center}
    \hspace*{1.0cm}\includegraphics[width=0.8\linewidth]{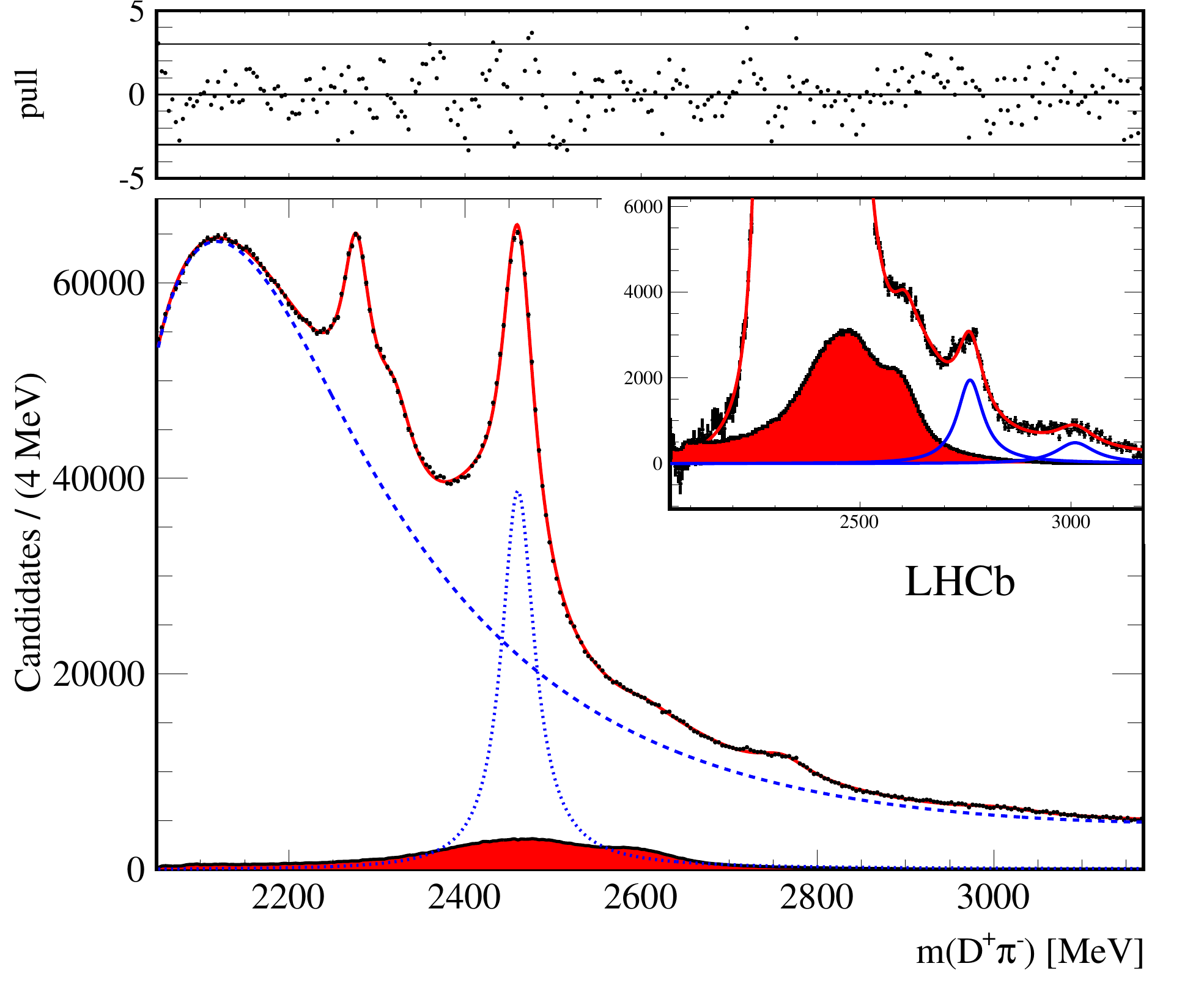}
    \vspace*{-0.5cm}
  \end{center}
  \caption{
    \small Fit to the $\Dp \pim$ mass spectrum. The filled histogram (in red) shows the estimated cross-feeds from the high mass $\Dstar \pi$ resonances. The dashed (blue) line displays
the fitted background. The dotted (blue) line shows the \DTwentyFourSixtyNeutral contribution. The inset displays the mass spectrum after the fitted background subtraction. The full (blue) curves show the \DTwentySevenSixtyNeutral and \DThreeNeutral contributions. The top window displays the pull distribution where the horizontal lines indicate $\pm 3 \sigma$.
    }
  \label{fig:fig11}
\end{figure}

Similarly, to obtain the expected lineshape of the cross-feed in the $\Dz \pip$ final state, we generate the four resonances according to the decays shown in Eq.~(\ref{dz_pip}). We assume, for the charged modes, rates for the four states similar to that for the neutral modes. The overall resulting structures obtained for the $D^{*0} \to D^0 \pi^0$ and $D^{*0} \to D^0 \gamma$ decays are scaled according to their branching
fractions and the distribution is scaled by the factor $R_{\Dz \pi^+}$ discussed above. The resulting contribution is superimposed on the $\Dz \pip$ mass spectrum shown in Fig.~\ref{fig:fig12}.
\begin{figure}[t]
  \begin{center}
    \hspace*{1.0cm}\includegraphics[width=0.8\linewidth]{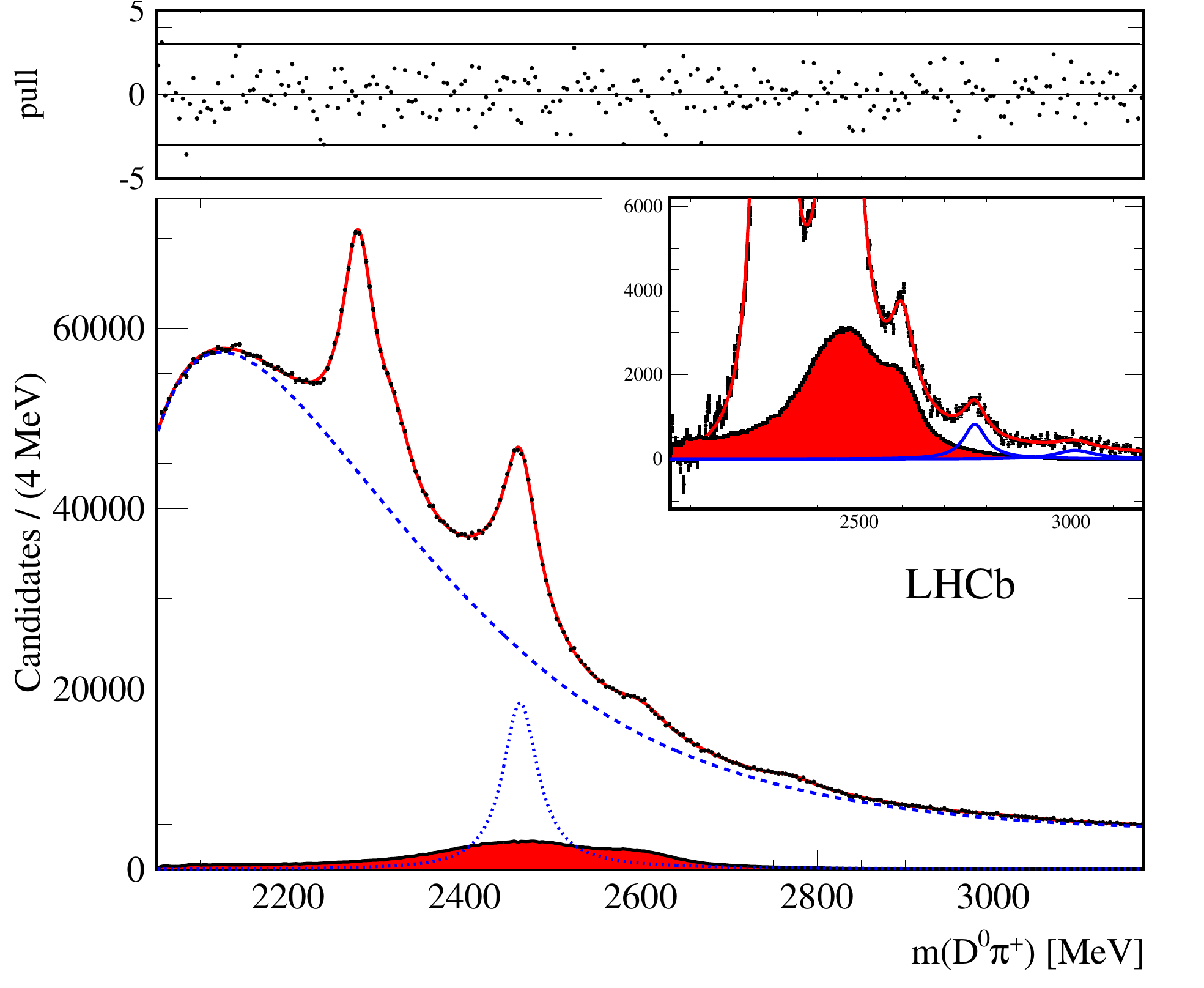}
    \vspace*{-0.5cm}
  \end{center}
  \caption{
    \small Fit to the $\Dz \pip$ mass spectrum. The filled histogram (in red) shows the estimated cross-feeds from the high mass $\Dstar \pi$ resonances. The dashed (blue) line displays
the fitted background. The dotted (blue) line shows the \DTwentyFourSixtyCharged contribution. The inset displays the mass spectrum after the fitted background subtraction. The full (blue) curves show the \DTwentySevenSixtyCharged and \DThreeCharged contributions. The top window displays the pull distribution where the horizontal lines indicate $\pm 3 \sigma$.
    }
  \label{fig:fig12}
\end{figure}

The cross-feed lineshapes obtained by the generator level simulation are not precise enough to be included in the fits to the $\Dp \pim$ and $\Dz \pip$ mass spectra.
We therefore follow an empirical procedure to obtain good fits in this mass region. We first notice that these contributions
produce a distortion of the \DTwentyFourSixtyNeutral and \DTwentyFourSixtyCharged lineshapes. These are accommodated in the fit by means
of a Breit-Wigner function, which we include to obtain a good description of the data. The parameters of the Breit-Wigner function are $M=2414.3\pm1.4$~\mev and $\Gamma=103.2 \pm 2.7$~\mev for the $\Dp \pim$ final state and $M=2435.1 \pm 5.2$~\mev and $\Gamma=106.9 \pm 6.2$~\mev for the $\Dz \pip$ final state. We consider these contributions as methods to improve the description of the cross-feeds.

We expect, in both the $\Dp \pim$ and $\Dz \pip$  mass spectra, the presence of \DTwentySixHundred and \DTwentySevenSixty states. Enhancements in these mass regions can be seen in the two mass spectra shown in Fig.~\ref{fig:fig11} and Fig.~\ref{fig:fig12}. However the \DTwentySixHundred region is strongly affected by cross-feed. We include a simple Breit-Wigner function to describe
these mass regions and obtain $M=2621.7 \pm 1.4$~\mev and $\Gamma=119.7 \pm 6.5$~\mev for the $\Dp \pim$ final state and $M=2599.9 \pm 0.9$~\mev and $\Gamma=72.3\pm 4.0$~\mev for the $\Dz \pip$ final state. However the parameters so far obtained are strongly biased by the presence of the cross-feed and we
therefore report, for the \DTwentySixHundred resonance, only the results obtained from the fit to the $\Dstarp \pim$ mass spectrum.

To obtain good quality fits we add broad structures around 3000~\mev, which we label \DThreeNeutral and \DThreeCharged. Their parameters are derived from the fit to the $\Dp \pim$ mass spectrum and then
fixed in the fit to the $\Dz \pip$ mass spectrum, where the effect is weaker.

The sensitivity of the fits to the presence of the broad \DTwentyFourHundred resonance is tested by performing simulations that include a \DTwentyFourHundred resonance with parameters fixed to their known
values and vary the background lineshape within a wide range of values. We find a high correlation between the \DTwentyFourHundred parameters and the background lineshape and a failure of the fit to
obtain correct estimates of its parameters and yields. Therefore this contribution is not included in the fit.

The fits to the $\Dp \pim$ and $\Dz \pip$ mass spectra are shown in Fig.~\ref{fig:fig11} and Fig.~\ref{fig:fig12}, respectively.
Masses, widths, yields, and significances for the all the fitted resonances are  displayed in Table~\ref{tab:fits}. All the statistical significances are well above 5$\sigma$.

\vspace{0.5cm}
\section{Cross-checks and systematic uncertainties}
\label{sec:sys}

Several cross-checks are performed to test the stability of the fits and their correct statistical behaviour.
We first repeat all the fits, including the spin-parity analysis, lowering the \pt requirement from 7.5 to 7.0 GeV. We find that all the resonance parameters vary within their statistical uncertainties and that the spin-parity assignments are not affected by this selection. 

The fits stability and the uncertainties on the resonance parameters are tested using random variations of the histogram contents. For each histogram, we 
obtain and fit 500 new histograms by random Poisson
variation of each bin content. We find in all cases a Gaussian behaviour of all the fit components with r.m.s. values that agree well with the statistical uncertainties given by the fits.

The systematic uncertainties on the resonance parameters and yields reported in Table~\ref{tab:fits} are estimated as follows.
The background lineshape uncertainty is estimated using an alternative function $B(m)=(m-m_{\rm th})^{a}e^{-b_1m - b_2m^2 - b_3m^3}$, where $m_{\rm th}$ is the threshold mass. 
This function gives acceptable fits for the $D \pi$ mass spectra but generally a worse description of the threshold region.

The background lineshapes are additionaly tested by random variation of their parameters. For each mass spectrum, we generate and fit 500 new histograms where the resonance parameters and 
yields are fixed to the values obtained from the data, while the background yield is fixed but has parameters varying within $\pm 3 \sigma$ from the values obtained from the data. The distributions obtained from these fits are used to obtain systematic uncertainties due to the background lineshape. 
For the uncertainty due to the background lineshape the largest value between the estimates from the two methods described above is taken. 

In the fits to the $\Dstarp \pim$ mass spectra, where resonances have in some cases fixed parameters, we let the resonance parameters float sequentially. The procedure is repeated for each
helicity sample and for the fit to the total mass spectrum.
The Breit-Wigner shapes used to describe the \DTwentySevenSixty resonance in the $\Dp \pim$ and  $\Dz \pip$ mass spectra are replaced by a relativistic Breit-Wigner functions with different spin assignments. We also include the \DTwentyFourHundred resonance with parameters fixed to the known values and obtain a small improvement in the fit to the $\Dp \pim$ mass spectrum but a yield consistent with zero in the fit to the $\Dz \pip$ mass spectrum.

The various estimated systematic uncertainties are added in quadrature.
We do not report systematic uncertainties on the structures labelled as \DThreeU and \DThreeCharged because, being at the limit of the mass spectra, they are strongly correlated with the background parameters.

\section{Discussion and conclusions}

A study of the $\Dp \pim$, $\Dz \pip$, and $\Dstarp \pim$ final states is reported using a sample of $pp$ collision data, corresponding to an integrated luminosity of 1.0\invfb, collected at a centre-of-mass energy of $7\tev$ with the \lhcb detector. We observe the \DTwentyFourTwentyNeutral resonance in the $\Dstarp \pim$ final state, and the \DTwentyFourSixty 
resonance in the $\Dp \pim$, $\Dz \pip$ and $\Dstarp \pim$ final states, measuring their parameters and confirming their spin-parity assignment~\cite{Beringer:1900zz}.
We also observe two natural parity resonances \DTwentySixHundredNeutral and \DTwentySevenSixtyNeutral in the $\Dstarp \pim$ mass spectrum and measure their angular distributions. The analysis of the $\Dp \pim$ and $\Dz \pip$ mass spectra supports the presence of \DTwentySevenSixty while the analysis of the \DTwentySixHundred region is inconclusive 
due to the presence of cross-feed from the resonances appearing in the $\Dstar \pi$ final state. 
The analysis of the $\Dstarp \pim$ final state also shows the presence of two unnatural parity states, \DTwentyFiveFiftyNeutral and \DTwentySevenFiftyNeutral, for which we
also perform a spin-parity analysis.
 
We observe a further structure in the $\Dstarp \pim$ final state, labelled as \DThreeU with an angular distribution that is compatible with unnatural parity. We also observe 
structures in the $\Dp \pim$ and $\Dz \pip$ mass spectra that we label as \DThreeNeutral and \DThreeCharged. The properties of all these structures are uncertain and could be
the result of a superposition of several 1F states, as expected by the quark model predictions~\cite{Godfrey:1985xj}.
The overall results from this analysis are in partial agreement with the results from \babar \ experiment~\cite{delAmoSanchez:2010vq}, although for some resonances, especially the \DTwentySixHundredNeutral state, we measure different parameters.

The main source of the difference between the two analyses is related to the method of obtaining the \DTwentySixHundredNeutral parameters which, in the \babar approach, are extracted
from the fit to the  $\Dp \pim$ mass spectrum and then fixed in the analysis of the $\Dstarp \pim$ mass spectrum. Due to the correlation between the resonances parameters, this procedure
also affects the properties of the other states appearing in the $\Dstarp \pim$ mass spectrum. In the present analysis, as stated above, we measure important cross-feeds in the 2500-2600~\mev region of the $\Dp \pim$ and $\Dz \pip$ final states and therefore we obtain the \DTwentySixHundredNeutral parameters from the $\Dstarp \pim$ final state only.

We compare the quark-model predictions given in Fig.~\ref{fig:Godfrey} with our mass measurements and spin-parity analysis reported in Table~\ref{tab:fits} and
Table~\ref{tab:spin}, respectively.
The \babar analysis suggests a $J^P=0^-$ assignment for the \DTwentyFiveFiftyNeutral state (labelled 2S $D_0(2558)$ in Fig.~\ref{fig:Godfrey}). Our results are consistent with the \babar \ measurement, but cannot confirm it, due to the superposition of many relatively broad resonances in a limited mass region which complicates the extraction of the resonances parameters.

The \DTwentySixHundredNeutral resonance is observed to decay to $\Dstarp \pim$, has natural parity and therefore is expected to decay to $\D \pi$.
However the presence of this state in the $\D \pi$ mass spectra is obscured by the presence of  cross-feeds from the $\Dstar \pi$ channels.
We tentatively identify the \DTwentySixHundredNeutral resonance as a $J^P=1^-$ state (2S $D^*_1(2618)$).

The \DTwentySevenSixtyNeutral is observed in the  $\Dstarp \pim$ and $\Dp \pim$ decay modes with consistent parameters. We also observe the \DTwentySevenSixtyCharged
in the $\Dz \pip$ final state which can be identified as a $J^P=1^-$ state (1D $D^*_1(2796)$).
The \DTwentySevenFiftyNeutral could be identified as the $J^P=2^-$ (1D $D_2(2801)$) resonance, although in this case the measured and predicted mass do not agree well. Definitive spin-parity assignments will be possible if these states are observed in $B$ decays.

\input{acknowledgements}

\addcontentsline{toc}{section}{References}
\bibliographystyle{LHCb}
\bibliography{main}

\end{document}

%% file: lhcb-symbols-def.tex
\def\lhcb {\mbox{LHCb}\xspace}
\def\ux85 {\mbox{UX85}\xspace}

\def\babar  {\mbox{BaBar}\xspace}
\def\belle  {\mbox{Belle}\xspace}

\ifthenelse{\boolean{uprightparticles}}%
{

 \def\Ppi         {\ensuremath{\uppi}\xspace}

 \def\PDelta      {\ensuremath{\Delta}\xspace}                 
 \def\PXi      {\ensuremath{\Xi}\xspace}                 
 \def\PLambda      {\ensuremath{\Lambda}\xspace}                 
 \def\PSigma      {\ensuremath{\Sigma}\xspace}                 
 \def\POmega      {\ensuremath{\Omega}\xspace}                 
 \def\PUpsilon      {\ensuremath{\Upsilon}\xspace}                 
 
 \def\PB      {\ensuremath{\mathrm{B}}\xspace}                 
                  
 \def\PD      {\ensuremath{\mathrm{D}}\xspace}

 \def\PK      {\ensuremath{\mathrm{K}}\xspace}

 \def\Pb      {\ensuremath{\mathrm{b}}\xspace}                 
 \def\Pc      {\ensuremath{\mathrm{c}}\xspace}

 \def\Pi      {\ensuremath{\mathrm{i}}\xspace}

}
{

 \def\Ppi         {\ensuremath{\pi}\xspace}

 \mathchardef\PDelta="7101
 \mathchardef\PXi="7104
 \mathchardef\PLambda="7103
 \mathchardef\PSigma="7106
 \mathchardef\POmega="710A
 \mathchardef\PUpsilon="7107
                  
 \def\PB      {\ensuremath{B}\xspace}                 
                  
 \def\PD      {\ensuremath{D}\xspace}

 \def\PK      {\ensuremath{K}\xspace}

 \def\Pb      {\ensuremath{b}\xspace}                 
 \def\Pc      {\ensuremath{c}\xspace}

 \def\Pi      {\ensuremath{i}\xspace}

}

\def\epem       {\ensuremath{\Pe^+\Pe^-}\xspace}

\def\cquark    {\ensuremath{\Pc}\xspace}

\def\bquark    {\ensuremath{\Pb}\xspace}

\def\pion  {\ensuremath{\Ppi}\xspace}
\def\piz   {\ensuremath{\pion^0}\xspace}

\def\pip   {\ensuremath{\pion^+}\xspace}
\def\pim   {\ensuremath{\pion^-}\xspace}

\def\kaon  {\ensuremath{\PK}\xspace}
  \def\Kbar  {\kern 0.2em\overline{\kern -0.2em \PK}{}\xspace}

\def\Kz    {\ensuremath{\kaon^0}\xspace}
\def\Kzb   {\ensuremath{\Kbar^0}\xspace}
\def\KzKzb {\ensuremath{\Kz \kern -0.16em \Kzb}\xspace}
\def\Kp    {\ensuremath{\kaon^+}\xspace}
\def\Km    {\ensuremath{\kaon^-}\xspace}

\def\KpKm  {\ensuremath{\Kp \kern -0.16em \Km}\xspace}

  \def\Dbar    {\kern 0.2em\overline{\kern -0.2em \PD}{}\xspace}
\def\D       {\ensuremath{\PD}\xspace}

\def\Dz      {\ensuremath{\D^0}\xspace}
\def\Dzb     {\ensuremath{\Dbar^0}\xspace}
\def\DzDzb   {\ensuremath{\Dz {\kern -0.16em \Dzb}}\xspace}
\def\Dp      {\ensuremath{\D^+}\xspace}
\def\Dm      {\ensuremath{\D^-}\xspace}

\def\DpDm    {\ensuremath{\Dp {\kern -0.16em \Dm}}\xspace}
\def\Dstar   {\ensuremath{\D^*}\xspace}

\def\Dstarp  {\ensuremath{\D^{*+}}\xspace}

\def\B       {\ensuremath{\PB}\xspace}
  \def\Bbar    {\kern 0.18em\overline{\kern -0.18em \PB}{}\xspace}

  \def\Y#1S{\ensuremath{\PUpsilon{(#1S)}}\xspace}

\def\Lbar {\ensuremath{\kern 0.1em\overline{\kern -0.1em\Lambda\kern -0.05em}\kern 0.05em{}}\xspace}

\def\to                 {\ensuremath{\rightarrow}\xspace}

\def\AT#1     {\ensuremath{A_{\mathrm{T}}^{#1}}\xspace}           

\def\C#1      {\ensuremath{\mathcal{C}_{#1}}\xspace}                       
\def\Cp#1     {\ensuremath{\mathcal{C}_{#1}^{'}}\xspace}                    
\def\Ceff#1   {\ensuremath{\mathcal{C}_{#1}^{\mathrm{(eff)}}}\xspace}        
\def\Cpeff#1  {\ensuremath{\mathcal{C}_{#1}^{'\mathrm{(eff)}}}\xspace}       
\def\Ope#1    {\ensuremath{\mathcal{O}_{#1}}\xspace}                       
\def\Opep#1   {\ensuremath{\mathcal{O}_{#1}^{'}}\xspace}                    

\newcommand{\tev}{\ensuremath{\mathrm{\,Te\kern -0.1em V}}\xspace}
\newcommand{\gev}{\ensuremath{\mathrm{\,Ge\kern -0.1em V}}\xspace}
\newcommand{\mev}{\ensuremath{\mathrm{\,Me\kern -0.1em V}}\xspace}
\newcommand{\kev}{\ensuremath{\mathrm{\,ke\kern -0.1em V}}\xspace}
\newcommand{\ev}{\ensuremath{\mathrm{\,e\kern -0.1em V}}\xspace}
\newcommand{\gevc}{\ensuremath{{\mathrm{\,Ge\kern -0.1em V\!/}c}}\xspace}
\newcommand{\mevc}{\ensuremath{{\mathrm{\,Me\kern -0.1em V\!/}c}}\xspace}
\newcommand{\gevcc}{\ensuremath{{\mathrm{\,Ge\kern -0.1em V\!/}c^2}}\xspace}
\newcommand{\gevgevcccc}{\ensuremath{{\mathrm{\,Ge\kern -0.1em V^2\!/}c^4}}\xspace}
\newcommand{\mevcc}{\ensuremath{{\mathrm{\,Me\kern -0.1em V\!/}c^2}}\xspace}

\def\mm   {\ensuremath{\rm \,mm}\xspace}

\def\mum  {\ensuremath{\,\upmu\rm m}\xspace}

\def\invfb   {\ensuremath{\mbox{\,fb}^{-1}}\xspace}

\def\gsim{{~\raise.15em\hbox{$>$}\kern-.85em
          \lower.35em\hbox{$\sim$}~}\xspace}
\def\lsim{{~\raise.15em\hbox{$<$}\kern-.85em
          \lower.35em\hbox{$\sim$}~}\xspace}

\def\pt         {\mbox{$p_{\rm T}$}\xspace}

\def\evtgen     {\mbox{\textsc{EvtGen}}\xspace}
\def\pythia     {\mbox{\textsc{Pythia}}\xspace}

\def\geant      {\mbox{\textsc{Geant4}}\xspace}

\def\tell1  {TELL1\xspace}
\def\ukl1   {UKL1\xspace}



%% file: title-LHCb-PAPER.tex

\begin{titlepage}
\pagenumbering{roman}

\vspace*{-1.5cm}
\centerline{\large EUROPEAN ORGANIZATION FOR NUCLEAR RESEARCH (CERN)}
\vspace*{1.5cm}
\hspace*{-0.5cm}
\begin{tabular*}{\linewidth}{lc@{\extracolsep{\fill}}r}
\ifthenelse{\boolean{pdflatex}}
{\vspace*{-2.7cm}\mbox{\!\!\!\includegraphics[width=.14\textwidth]{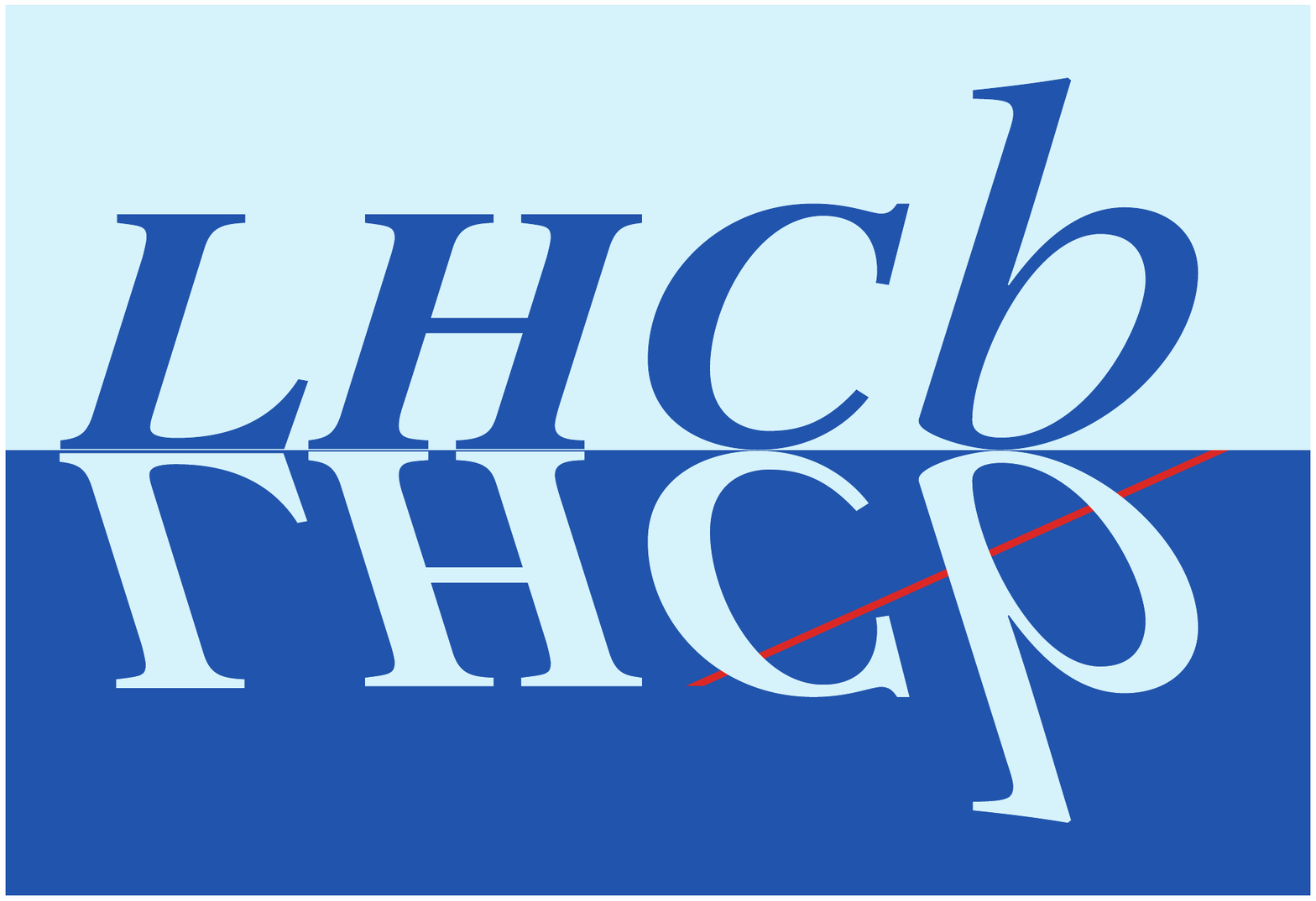}} & &}%
{\vspace*{-1.2cm}\mbox{\!\!\!\includegraphics[width=.12\textwidth]{lhcb-logo.eps}} & &}%
\\
 & & CERN-PH-EP-2013-122 \\  
 & & LHCb-PAPER-2013-026 \\  
 & & 17 July 2013 \\ 
 & & \\
\end{tabular*}

\vspace*{3.0cm}

{\bf\boldmath\huge
\begin{center}
Study of $\DJ$ meson decays to \DPi, \DzPi and \DstarPi final states in $pp$ collisions
\end{center}
}

\vspace*{1.5cm}

\begin{center}
The LHCb collaboration\footnote{Authors are listed on the following pages.}
\end{center}

\vspace{\fill}

\begin{abstract}
  \noindent
A study of \DPi, \DzPi and \DstarPi final states is performed using $pp$ collision data, corresponding to an integrated luminosity of 1.0\invfb, 
collected at a centre-of-mass energy of $7\tev$ with the \lhcb detector. 
The \DTwentyFourTwentyNeutral resonance is observed in the  \DstarPi final state and the \DTwentyFourSixty resonance is observed in the 
\DPi, \DzPi and \DstarPi final states. For both resonances, their properties and spin-parity assignments are obtained.
In addition, two natural parity and two unnatural parity resonances are observed in the mass region between 2500 and 2800~\mev.
Further structures in the region around 3000 \mev are observed in all the \DstarPi, \DPi and \DzPi final states.
\end{abstract}

\vspace*{2.0cm}

\begin{center}
Submitted to JHEP 
\end{center}

\vspace{\fill}

{\footnotesize 
\centerline{\copyright~CERN on behalf of the \lhcb collaboration, license \href{http://creativecommons.org/licenses/by/3.0/}{CC-BY-3.0}.}}
\vspace*{2mm}

\end{titlepage}


\newpage
\setcounter{page}{2}
\mbox{~}
\newpage

\input{LHCb_authorlist}

\cleardoublepage

%% file: LHCb_authorlist.tex
\centerline{\large\bf LHCb collaboration}
\begin{flushleft}
\small
R.~Aaij$^{40}$, 
B.~Adeva$^{36}$, 
M.~Adinolfi$^{45}$, 
C.~Adrover$^{6}$, 
A.~Affolder$^{51}$, 
Z.~Ajaltouni$^{5}$, 
J.~Albrecht$^{9}$, 
F.~Alessio$^{37}$, 
M.~Alexander$^{50}$, 
S.~Ali$^{40}$
% ...

{\footnotesize \it
% ...
$ ^{5}$Clermont Universit\'{e}, Universit\'{e} Blaise Pascal, CNRS/IN2P3, LPC, Clermont-Ferrand, France\\
$ ^{6}$CPPM, Aix-Marseille Universit\'{e}, CNRS/IN2P3, Marseille, France\\
% ...
$ ^{9}$Fakult\"{a}t Physik, Technische Universit\"{a}t Dortmund, Dortmund, Germany\\
% ...
$ ^{36}$Universidad de Santiago de Compostela, Santiago de Compostela, Spain\\
$ ^{37}$European Organization for Nuclear Research (CERN), Geneva, Switzerland\\
% ...
$ ^{40}$Nikhef National Institute for Subatomic Physics, Amsterdam, The Netherlands\\
% ...
$ ^{45}$H.H. Wills Physics Laboratory, University of Bristol, Bristol, United Kingdom\\
% ...
$ ^{50}$School of Physics and Astronomy, University of Glasgow, Glasgow, United Kingdom\\
$ ^{51}$Oliver Lodge Laboratory, University of Liverpool, Liverpool, United Kingdom\\
% ...
}
\end{flushleft}

%% file: acknowledgements.tex
\section*{Acknowledgements}

\noindent We express our gratitude to our colleagues in the CERN
accelerator departments for the excellent performance of the LHC. We
thank the technical and administrative staff at the LHCb
institutes. We acknowledge support from CERN and from the national
agencies: CAPES, CNPq, FAPERJ and FINEP (Brazil); NSFC (China);
CNRS/IN2P3 and Region Auvergne (France); BMBF, DFG, HGF and MPG
(Germany); SFI (Ireland); INFN (Italy); FOM and NWO (The Netherlands);
SCSR (Poland); MEN/IFA (Romania); MinES, Rosatom, RFBR and NRC
``Kurchatov Institute'' (Russia); MinECo, XuntaGal and GENCAT (Spain);
SNSF and SER (Switzerland); NAS Ukraine (Ukraine); STFC (United
Kingdom); NSF (USA). We also acknowledge the support received from the
ERC under FP7. The Tier1 computing centres are supported by IN2P3
(France), KIT and BMBF (Germany), INFN (Italy), NWO and SURF (The
Netherlands), PIC (Spain), GridPP (United Kingdom). We are thankful
for the computing resources put at our disposal by Yandex LLC
(Russia), as well as to the communities behind the multiple open
source software packages that we depend on.